\documentclass[lettersize,journal]{IEEEtran}
\usepackage{amsmath,amsfonts}
\usepackage{cases}
\usepackage[ruled]{algorithm2e} 
\usepackage{array}
\usepackage[caption=false,font=normalsize,labelfont=sf,textfont=sf]{subfig}
\usepackage{textcomp}
\usepackage{stfloats}
\usepackage{url}
\usepackage{verbatim}
\usepackage{graphicx}
\usepackage{cite}
\usepackage{CJKutf8}
\usepackage{amssymb}
\newtheorem{remark}{Remark}
\usepackage{xspace}
\usepackage{float}

\begin{document}
\begin{CJK}{UTF8}{gbsn}
\title{Distributed fusion filter over lossy wireless sensor networks with the presence of non-Gaussian noise (This work has been submitted to the
IEEE for possible publication. Copyright may be
transferred without notice, after which this version
may no longer be accessible.)}

\author{Jiacheng He, Bei Peng, Zhenyu Feng, Xuemei Mao, Song Gao, Gang Wang
\thanks{This study was founded by the National Natural Science Foundation of China with Grant 51975107 and Sichuan Science and Technology Major Project No.2022ZDZX0039, No.2019ZDZX0020，and Sichuan Science and Technology Program No.2022YFG0343.}
\thanks{Manuscript received April 19, 2021; revised August 16, 2021.}}

\markboth{Journal of \LaTeX\ Class Files,~Vol.~14, No.~8, August~2021}%
{Shell \MakeLowercase{\textit{et al.}}: A Sample Article Using IEEEtran.cls for IEEE Journals}

\maketitle

\begin{abstract}
The information transmission between nodes in a wireless sensor networks (WSNs) often causes packet loss due to denial-of-service (DoS) attack, energy limitations, and environmental factors, and the information that is successfully transmitted can also be contaminated by non-Gaussian noise. The presence of these two factors poses a challenge for distributed state estimation (DSE) over WSNs. In this paper, a generalized packet drop model is proposed to describe the packet loss phenomenon caused by DoS attacks and other factors. Moreover, a modified maximum correntropy Kalman filter is given, and it is extended to distributed form (DM-MCKF). In addition, a distributed modified maximum correntropy Kalman filter incorporating the generalized data packet drop (DM-MCKF-DPD) algorithm is provided to implement DSE with the presence of both non-Gaussian noise pollution and packet drop. A sufficient condition to ensure the convergence of the fixed-point iterative process of the DM-MCKF-DPD algorithm is presented and the computational complexity of the DM-MCKF-DPD algorithm is analyzed. Finally, the effectiveness and feasibility of the proposed algorithms are verified by simulations.
\end{abstract}


\section{Introduction}
Recently, the state estimation in wireless sensor networks (WSNs) has been a subject of considerable interest, and it has been applied widely, such as in static or dynamic target positioning \cite{9520133} and tracking \cite{8576615}, indoor positioning \cite{9031316}, vehicle navigation \cite{ZHANG2022111258}, and others \cite{7434638, HUANG2023293}. The most popular state estimation methods over WSNs are centralized and distributed fusion architectures. Distributed state estimation (DSE)  strategies with their strong robustness and low computational complexity have come to be known as the dominant state estimation algorithm for WSNs.

In distributed estimation methods, each node functions as a sensor and a fusion center, gathering data from itself and its neighbors to produce local estimation. However, information interaction with neighboring nodes often suffers from packet loss. The causes of packet loss usually include the following: 1) the sending of sensor information is usually randomly activated \cite{YANG20142070,liu2015event} or event-based \cite{9851668, HUANG20218754} due to long-term latent demand or energy limitations \cite{8114330}, the intermittent transmission of information can be considered a special case of packet loss; 2) the failure of information transmission due to the performance of the sensor itself or environmental factors \cite{guo2020intelligent}; 3) the DoS attacks in WSNs \cite{8653498}. These factors inevitably cause packet loss in the transmission of information between neighbors. The packet loss phenomenon is a potential source of instability, and it can lead to a further increase in the differences in node estimates in WSNs, which poses difficulties for the DSE of the network. Therefore, DSE that takes into account packet loss phenomena has gained significant attention.

To further reduce the difference in estimates between nodes due to packet loss, numerous research results have been carried out, and the consensus-based \cite{9257077} method and the diffusion-based \cite{5411741} method are two often employed techniques. A distributed Kalman filter (KF) \cite{7172030} based on a diffusion approach is developed with intermittent measurement, and the diffusion method is extended to nonlinear systems \cite{2018Diffusion}. Based on the above algorithms, a distributed diffusion unscented KF taking into account the unknown correlations in WSNs is derived \cite{CHEN2021109769}. Moreover, the discussion of consensus-based DSE over WSNs has also gained significant attention, some distributed Kalman consensus filters \cite{LI20153764,8409298} are developed over WSNs with intermittent observations for linear systems. For the DSE of nonlinear systems, the weighted average consensus-based cubature information filter \cite{8028780} is derived with the presence of measurement loss. In addition, from a certain point of view, stochastic sensor \cite{YANG20142070,liu2015event} or link \cite{YANG2017109} activation, or the DoS attacks \cite{7930412} can be considered as special cases of packet loss over WSNs. The study of different distributed fusion strategies constitutes an important branch of the lossy WSNs. On top of that, the multiplicative noises \cite{MA2017268,TIAN2016126} and correlated additive noises \cite{CABALLEROAGUILA201770,YAO2023108829,DING2019138}, and these algorithms use the random parameter measurement matrices. 

The above-mentioned studies take into account that the measurement being successfully passed between nodes contains Gaussian noise, and they perform well in Gaussian environments. However, due to the fact that non-Gaussian noise is common in practical application \cite{HE2022,8540327,WANG2019115,9923771,HE2023108787}, which results in measurements that are successfully delivered being often contaminated by non-Gaussian noise, and this situation inevitably degrades the performance of the algorithm under Gaussian assumption. Several studies have focused on conducting state estimation in WSNs under a non-Gaussian noise environment. Distributed particle filters (DPFs) \cite{gu2007distributed} have been developed for nonlinear and non-Gaussian noise systems, however, the high computational complexity of DPF has always been the main factor limiting its wide application. A new idea, which introduces the concept of information theoretic learning \cite{5952087} as a new cost function, has been widely employed to create a new distributed Kalman filter algorithm in recent years. The distributed maximum correntropy KF (DMCKF) algorithm \cite{wang2019distributed} and its variants \cite{hu2022efficient,wang2021distributed} are developed to reduce the influence of non-Gaussian noise. To further enhance the performance of the DSE under non-Gaussian conditions, the distributed minimum error entropy KF (DMEEKF) algorithm is developed in \cite{feng2023distributed}. Due to the double summation of minimum error entropy criteria, the computational complexity of the DMEEKF algorithm is higher than the DMCKF algorithm, and it is a burden for sensor nodes with energy limitations. Obviously, it is still an open and nontrivial work to investigate the DSE issues over lossy WSNs with measurement contaminated by non-Gaussian noise, which also constitutes the main motivation for this article.

In this paper, we first propose a generalized packet loss model to describe the process of packet loss due to energy constraints and DoS attacks. A modified maximum correntropy Kalman filter (M-MCKF) is proposed on top of the analysis of the traditional KF algorithm. Moreover, the M-MCKF is extended to a distributed form, and one can get the distributed M-MCKF (DM-MCKF) algorithm. In addition, a distributed M-MCKF algorithm incorporating the proposed generalized packet loss model is called the DM-MCKF-DPD algorithm. Moreover, a sufficient condition is provided that can ensure the convergence of the DM-MCKF-DPD algorithm's fixed-point iterations.

The rest of this work is structured as follows. In Section \ref{section:Preliminaries}, the problem formulation is briefly reviewed. In Section \ref{section:derivation}, the derivation, computational complexity, and convergence issue of the DM-MCKF-DPD algorithm are presented. The simulation examples are provided in Section \ref{section:simulations}, and, finally, the conclusion is given in Section \ref{section:Conclusion}.

\section{Problem formulation} \label{section:Preliminaries}
\subsection{Distributed Kalman filter}
Consider a WSNs with ${N}$ sensors, and the states and observations are as follows:
\begin{equation}
\begin{split}
\left\{ \begin{gathered}
  {{\boldsymbol{x}}_k} = {{\boldsymbol{A}}_k}{{\boldsymbol{x}}_{k - 1}} + {{\boldsymbol{q}}_{k - 1}}, \hfill \\
  {\boldsymbol{y}}_k^i = {{\boldsymbol{C}}^i}{{\boldsymbol{x}}_k} + {\boldsymbol{v}}_k^i,{\text{  }}i = 1,2, \cdots ,N,{\text{and }}i \in \Omega , \hfill \\ 
\end{gathered}  \right.
\end{split}
\end{equation}

where ${{{\boldsymbol{x}}_{k - 1}} \in {\mathbb{R}^{n \times 1}}}$ represents the states of the dynamical system at instants ${k - 1}$, ${\Omega }$ represents the set of all sensors in the network, ${\boldsymbol{y}}_k^i \in {\mathbb{R}^{{m_i} \times 1}}({m_i} = \operatorname{rank} ({{\boldsymbol{C}}_i}),\forall i \in \Omega )$ denotes the observations obtained by node ${i}$ at instant ${k}$, ${{{\boldsymbol{A}}_k}}$ and ${{{\boldsymbol{C}}^i}}$ represent the state transition matrix and observation matrix of the system, respectively, and ${{{\boldsymbol{q}}_k}}$ and ${{\boldsymbol{v}}_k^i}$ are the mutually uncorrelated process noise and measurement noise, respectively, with zero means and covariances ${{\boldsymbol{Q}}_{k - 1}}$ and ${\boldsymbol{R}}_{k}^i$. We then define ${{\Omega _i} \subset \Omega }$ to represent the set of all neighboring sensors of the ${i}$th sensor. Therefore, set ${{\mho _i} = {\Omega _i} \cup \{ i\}}$ contains node ${i}$ itself and all its neighboring nodes. The main steps of DKF include state prediction and update.

1) state prediction: use the state ${\boldsymbol{\hat x}}_{k - 1|k - 1}^i$ and covariance ${\boldsymbol{P}}_{k - 1|k - 1}^i$ of the $k-1$ time to predict the state ${\boldsymbol{\hat x}}_{k|k - 1}^i$ and covariance ${\boldsymbol{P}}_{k|k - 1}^i$, and the specific methods are presented in
\begin{align}\label{akkj1dAkJJ}
{\boldsymbol{\hat x}}_{k|k - 1}^i = {{\boldsymbol{A}}_k}{\boldsymbol{\hat x}}_{k - 1|k - 1}^i
\end{align}
and
\begin{align}\label{akk1234}
{\boldsymbol{P}}_{k|k - 1}^i = {{\boldsymbol{A}}_k}{\boldsymbol{P}}_{k - 1|k - 1}^i{\boldsymbol{A}}_k^T + {{\boldsymbol{Q}}_{k - 1}},
\end{align}
where ${\left(  \cdot  \right)^T}$ is the transpose operation.

2) state update: update the state ${\boldsymbol{\hat x}}_{k|k}^i$ and covariance ${\boldsymbol{P}}_{k|k}^i$ using measurement information ${\boldsymbol{y}}_k^{{\mho _i}} = {\text{vec}}{\{ {\boldsymbol{y}}_k^j\} _{j \in {\mho _i}}}$ and gain ${\boldsymbol{K}}_k^i$, and the specific steps are
\begin{align}\label{KkidyPkkiCJ}
{\boldsymbol{K}}_k^i = {\boldsymbol{P}}_{k|k - 1}^i{\text{ }}{\left( {{{\boldsymbol{C}}^{{\mho _i}}}} \right)^T}{\left[ {{{\boldsymbol{C}}^{{\mho _i}}}{\boldsymbol{P}}_{k|k - 1}^i{{\left( {{{\boldsymbol{C}}^{{\mho _i}}}} \right)}^T} + {\boldsymbol{R}}_k^{{\mho _i}}} \right]^{ - 1}},
\end{align}
\begin{align}\label{xkkijKki}
{\boldsymbol{\hat x}}_{k|k}^i = {\boldsymbol{\hat x}}_{k|k - 1}^i + {\boldsymbol{K}}_k^i\left( {{\boldsymbol{y}}_k^{{\mho _i}} - {{\boldsymbol{C}}^{{\mho _i}}}{\boldsymbol{\hat x}}_{k|k - 1}^i} \right),
\end{align}
and
\begin{align}\label{PkkideyIjanKkiC}
{\boldsymbol{P}}_{k|k}^i = \left( {{\boldsymbol{I}} - {\boldsymbol{K}}_k^i{{\boldsymbol{C}}^{{\mho _i}}}} \right){\boldsymbol{P}}_{k|k - 1}^i,
\end{align}
where ${{\boldsymbol{C}}^{{\mho _i}}} = {\text{col}}{\{ {{\boldsymbol{C}}^j}\} _{j \in {\mho _i}}}$, and ${\text{col}}\left\{  \cdot  \right\}$ is the columnization operation; ${\boldsymbol{R}}_k^{{\mho _i}} = \operatorname{diag} {\{ {\boldsymbol{R}}_k^j\} _{j \in {\mho _i}}}$, $\operatorname{diag} \left\{  \cdot  \right\}$ is the block diagonalization operation.

From the above algorithm flow, the distributed Kalman filter method can be observed to operate by collecting measurements from nearby nodes and its own observations. However, the transmitted measurements from neighboring nodes often results in packet loss in wireless sensor networks due to intermittent sensor failures, stochastic communication link activation, and DoS attacks.

\subsection{Generalized packet drop model}
The state-space model of the DKF over the nodes of WSNs under conditions of data packet drops is the first problem that must be solved to estimate the state of a target node. The following model can be used to explain how node ${i}$ receives the measurements from its neighbouring node ${j}$ at instant ${k}$:
\begin{equation}
\begin{split}
{\boldsymbol{s}}_k^{i,j} \triangleq {\boldsymbol{f}}_k^{i,j}({\boldsymbol{y}}_k^j),{\text{  }}i \in \Omega ,{\text{  }}j \in {\mho _i},
\end{split}
\end{equation}
where ${\boldsymbol{s}}_k^{i,j}$ is the data obtained by node $i$ from node $j$. ${\boldsymbol{f}}_k^{i,j}\left(  \cdot  \right)$ is an arbitrary function, and it represents the mapping relationship between ${\boldsymbol{s}}_k^{i,j}$ and ${{\boldsymbol{y}}_k^j}$. When ${\boldsymbol{f}}_k^{i,j}\left(  \cdot  \right)$ is a nonlinear function, the function can be regarded as a nonlinear attack model \cite{7011170} leading to packet drop. For the linear case, ${\boldsymbol{f}}_k^{i,j}\left(  \cdot  \right)$ is a linear transformation of ${{\boldsymbol{y}}_k^j}$. The proposed linear model is defined as
\begin{align}\label{tkijjykj}
{\boldsymbol{s}}_k^{i,j} \triangleq {\boldsymbol{T}}_k^{i,j}{\boldsymbol{y}}_k^j,
\end{align}
where ${\boldsymbol{T}}_k^{i,j} \in {\mathbb{R}^{{m_j} \times {m_j}}}$ is an arbitrary matrix. \eqref{tkijjykj} can be used to represent linear attack strategies \cite{guo2016optimal,7930412}, and it can also be regarded as a stochastic sensor or link activation model \cite{YANG20142070,YANG2017109}, and event-based model \cite{liu2015event}. 
\begin{remark}
When DoS attack occurs, the measurements may not be delivered to the local fusion center. \eqref{tkijjykj} can be expressed as
\begin{align}\label{equ:iosjoosjkj}
{\boldsymbol{T}}_k^{i,j} = \gamma _k^{i,j}{{\boldsymbol{I}}_{{m_j}}},
\end{align}
\end{remark}
where the flag term $\gamma _k^{i,j}{\text{ = }}0$ or $\gamma _k^{i,j}{\text{ = }}1$ means DoS attack does not occur or occurred. More generally, flag term $\gamma _k^{i,j}$ are used to indicate the successful and unsuccessful transfer of information between nodes $i$ and $j$ at time $k$. Similarly, due to energy limitations, sensor link random activation and event-triggered based activation can be modeled as \eqref{equ:iosjoosjkj}, the flag term $\gamma _k^{i,j}{\text{ = }}0$ or $\gamma _k^{i,j}{\text{ = }}1$ represent inactive or active communication links respectively.

For this type of stochastic intermittent transmission of measurement ${{\boldsymbol{y}}_k^j}$ in lossy WSNs. It is assumed that the process has independent and identically distributed properties for ${\gamma _k^{i,j}}$ with the probability
\begin{equation}
\begin{split}
\operatorname{P} \{ \gamma _k^{i;j} = 1\}  = p_k^{i;j} > 0,{\text{  }}\forall k \geqslant 0,
\end{split}
\end{equation}
and $\operatorname{E} [\gamma _k^{i;j} = 1] = p_k^{i;j}$ and ${(\mu _k^{i;j})^2} = \operatorname{E} [{(\gamma _k^{i;j} - p_k^{i;j})^2}] = p_k^{i;j} - {(p_k^{i;j})^2}$.

It is assumed that variables ${\gamma _k^{i,j}}$ and ${\gamma _k^{j,i}}$ are mutually independent (i.e. ${\forall i \ne j}$), but the condition $p_k^{i,j} = p_k^{j,i}$ is allowed, and ${\gamma _k^{i,j}}$ is independent of the process noise, measurement noise, and initial state of the dynamical system. According to (\ref{equ:iosjoosjkj}), if sensor ${i}$ receives the observations from its neighboring node ${j}$ at instant ${k}$, then ${\gamma _k^{i,j} = 1}$ and ${\gamma _k^{i,j} = 0}$ when all the components of ${{{\boldsymbol{y}}_k^j}}$ are lost. We assume that node ${i}$ can receive all observations from itself at any time, so ${\gamma _k^{i,i} \equiv 1}$. 

According to the above discussion, the state-space model of the ${i}$th sensor in the case of data packet drops can be written as
\begin{equation}\label{equ:kiqj1jkisaa}
\begin{split}
{{\boldsymbol{x}}_k} = {\boldsymbol{A}}{{\boldsymbol{x}}_{k - 1}} + {{\boldsymbol{q}}_{k - 1}},
\end{split}
\end{equation}
\begin{equation}\label{equ:aa}
\begin{split}
{\boldsymbol{y}}_k^{{\mho _i}} = {{\boldsymbol{C}}^{{\mho _i}}}{\boldsymbol{x}}_k^i + {\boldsymbol{v}}_k^{{\mho _i}},
\end{split}
\end{equation}
and
\begin{equation}\label{equ:kiyikd}
\begin{split}
{\boldsymbol{s}}_k^{{\mho _i}} = {\boldsymbol{D}}_{\gamma ;k}^{{\mho _i}}{\boldsymbol{y}}_k^{{\mho _i}},
\end{split}
\end{equation}
where ${\boldsymbol{v}}_k^{{\mho _i}}{\text{ }} = {\text{vec}}{\{ {\boldsymbol{v}}_k^j\} _{j \in {\mho _i}}}$, ${\boldsymbol{s}}_k^{{\mho _i}} = {\text{vec}}{\{ {\boldsymbol{s}}_k^j\} _{j \in {\mho _i}}}$, and ${\boldsymbol{D}}_{\gamma ;k}^{{\mho _i}} = \operatorname{diag} {\{ \gamma _k^{i,j}{{\boldsymbol{I}}_{{m_j}}}\} _{j \in {\mho _i}}}$; ${{{\boldsymbol{I}}_{{m_j}}}(j \in {\mho _i})}$ represents an identity matrix of dimension ${{m_j} \times {m_j}}$. 
\subsection{Existing problem}
All the observations obtained by node $i$ are contained in ${\boldsymbol{s}}_k^{{\mho _i}}$. Due to packet loss, some of the measurements of neighboring nodes are not delivered to ${\boldsymbol{s}}_k^{{\mho _i}}$ smoothly. The measurements that are delivered to ${\boldsymbol{s}}_k^{{\mho _i}}$ are usually considered in the existing literature only when they are disturbed by Gaussian noise. However, non-Gaussian noise is very common in WSNs, which means that the matrix ${\boldsymbol{s}}_k^{{\mho _i}}$ is not only affected by communication packet loss, but also contaminated by non-Gaussian observation noise. In this paper, the measurements information matrix ${\boldsymbol{s}}_k^{{\mho _i}}$, which is affected by both communication packet loss and non-Gaussian noise, is used to perform DSE over WSNs. The generalized packet drop model is incorporated into the proposed distributed fusion filter, and a solution is provided that enables distributed state estimation under the coexistence of non-Gaussian noise and packet loss.

\section{Proposed DM-MCKF-DPD algorithm}\label{section:derivation}
\subsection{Correntropy}
The concept of correntropy, first proposed by Principe \emph{et al.}\cite{liu2007correntropy}, is a very practical method for evaluating the similarity between random variables $X,Y \in \mathbb{R}$ with the same dimensions. Here, correntropy is defined as
\begin{equation}
\begin{split}
V\left( {X,Y} \right) = \operatorname{E} \left[ {\kappa \left( {X,Y} \right)} \right] = \int {\kappa \left( {x,y} \right)} d{F_{XY}}\left( {x,y} \right),
\end{split}
\end{equation}
where ${{\text{E[}}{\text{.]}}}$ is the expectation operator, $V\left(  \cdot  \right)$ is the information potential, ${F_{XY}}\left( {x,y} \right)$ denotes the probability distribution function (PDF) with on ${X}$ and ${Y}$, and ${\kappa \left( { \cdot {\rm{ }},{\rm{ }} \cdot } \right)}$ is the shift-invariant Mercer kernel. Here, we employ the Gaussian kernel, which is given as
\begin{equation}\label{equ:epxeeg}
\begin{split}
\kappa \left( {x,y} \right) = {G_\sigma }\left( e \right) = \exp \left( { - \frac{{{e^2}}}{{2{\sigma ^2}}}} \right),
\end{split}
\end{equation}
where ${e = x - y}$ represents the error between elements ${x}$ and ${y}$, and ${\sigma  > 0}$ represents the kernelwidth (or kernel size) of the Gaussian kernel function.

However, only a limited amount of data related to the variables ${X}$ and ${Y}$ can be obtained in realistic scenarios, and the PDF ${F_{XY}}\left( {x,y} \right)$ is usually unknown. Under these conditions, a sample estimator can be utilized to calculate the correntropy as follows:
\begin{equation}
\begin{split}
\hat V\left( {X,Y} \right) = \frac{1}{L}\sum\limits_{l = 1}^L {{G_\sigma }} \left( {{e^l}} \right),
\end{split}
\end{equation}
where
\begin{equation}
\begin{split}
{e^l} = {x^l} - {y^l},\left( {{x^l},{y^l} \in \left\{ {{x^l},{y^l}} \right\}_{l = 1}^L} \right),
\end{split}
\end{equation}
and ${L}$ samples are employed to define ${{\operatorname{F} _{XY}}(x,y)}$. Compared with other similarity measurement schemes, such as the mean-square error (MSE) criterion, correntropy contains all even-order moments and is therefore useful for nonlinear and signal processing applications in non-Gaussian noise environments.
\subsection{The adjustment mechanisms of the KF and DKF}
According to the matrix inversion lemma \cite{chen2017maximum}, \eqref{PkkideyIjanKkiC} can be rewritten as
\begin{align}\label{PkkdenyuPkkiji1jcJ}
{\boldsymbol{P}}_{k|k}^i = {\left[ {{{\left( {{\boldsymbol{P}}_{k|k - 1}^i} \right)}^{ - 1}} + {{\left( {{{\boldsymbol{C}}^{{\mho _i}}}} \right)}^T}{{\left( {{\boldsymbol{R}}_k^{{\mho _i}}} \right)}^{ - 1}}{{\boldsymbol{C}}^{{\mho _i}}}} \right]^{ - 1}}.
\end{align}
Combining \eqref{KkidyPkkiCJ} and \eqref{PkkideyIjanKkiC}, one can get 
\begin{align}\label{KRdegyuPC}
{\boldsymbol{K}}_k^i{\boldsymbol{R}}_k^{{\mho _i}} = {\boldsymbol{P}}_{k|k}^i{\left( {{{\boldsymbol{C}}^{{\mho _i}}}} \right)^T}.
\end{align}
Substitute \eqref{PkkdenyuPkkiji1jcJ} and \eqref{KRdegyuPC} into \eqref{xkkijKki} yields 
\begin{align}\label{Apxbarikshukj1}
\begin{gathered}
  {\boldsymbol{\hat x}}_{k|k}^i = {\boldsymbol{\hat x}}_{k|k - 1}^i + {\boldsymbol{K}}_k^i\left( {{\boldsymbol{y}}_k^{{\mho _i}} - {{\boldsymbol{C}}^{{\mho _i}}}{\boldsymbol{\hat x}}_{k|k - 1}^i} \right) \hfill \\
   = {\boldsymbol{\hat x}}_{k|k - 1}^i - {\boldsymbol{K}}_k^i{{\boldsymbol{C}}^{{\mho _i}}}{\boldsymbol{\hat x}}_{k|k - 1}^i + {\boldsymbol{K}}_k^i{\boldsymbol{y}}_k^{{\mho _i}} \hfill \\
   = {{\boldsymbol{A}}_{\boldsymbol{P}}}{\boldsymbol{\hat x}}_{k|k - 1}^i + {{\boldsymbol{A}}_{\boldsymbol{R}}}{\boldsymbol{y}}_k^{{\mho _i}}, \hfill \\ 
\end{gathered} 
\end{align}
with
\begin{align}
\left\{ {\begin{array}{*{20}{l}}
  {{{\boldsymbol{A}}_{\boldsymbol{P}}} = {{\boldsymbol{M}}_k}{{\left( {{\boldsymbol{P}}_{k|k - 1}^i} \right)}^{ - 1}},} \\ 
  {{{\boldsymbol{A}}_{\boldsymbol{R}}} = {{\boldsymbol{M}}_k}{{\left( {{{\boldsymbol{C}}^{{\mho _i}}}} \right)}^T}{{\left( {{\boldsymbol{R}}_k^{{\mho _i}}} \right)}^{ - 1}},} 
\end{array}} \right.
\end{align}
and ${{\boldsymbol{M}}_k} = {[{({\boldsymbol{P}}_{k|k - 1}^i)^{ - 1}} + {({{\boldsymbol{C}}^{{\mho _i}}})^T}{({\boldsymbol{R}}_k^{{\mho _i}})^{ - 1}}{{\boldsymbol{C}}^{{\mho _i}}}]^{ - 1}}$.

From \eqref{Apxbarikshukj1}, the updated state ${\boldsymbol{\hat x}}_{k|k}^i$ can be regarded as the linear combination of the predicted state ${\boldsymbol{\hat x}}_{k|k - 1}^i$ and measurement ${\boldsymbol{y}}_k^{{J_i}}$. Matrices ${{{\boldsymbol{A}}_{\boldsymbol{P}}}}$ and ${{{\boldsymbol{A}}_{\boldsymbol{R}}}}$ are able to adjust the weights of ${\boldsymbol{\hat x}}_{k|k - 1}^i$ and ${\boldsymbol{y}}_k^{{J_i}}$ according to the statistical properties of the process noise and the measurement noise. Specifically, ${\boldsymbol{P}}_{k|k - 1}^i$ and ${{\boldsymbol{R}}_k^{{\mho _i}}}$ are central to the adjustment of weights, and ${\boldsymbol{P}}_{k|k - 1}^i$ is determined by ${{\boldsymbol{Q}}_{k - 1}}$.  The complete analysis of the adaptation mechanisms of weights ${{{\boldsymbol{A}}_{\boldsymbol{P}}}}$ and ${{{\boldsymbol{A}}_{\boldsymbol{R}}}}$ to the noise statistical properties of the algorithm is very complex and difficult. For the special case in which both the observation vector and the state vector are one-dimensional, which is also a case of KF. As the variance $R_k^{{\mho _i}}$ of the measurement noise becomes larger, the scalar ${A_R}$ will become smaller, which will reduce the proportion of the measurement $y_k^{{J_i}}$ in the updated state $\hat x_{k|k}^i$; a smaller variance of the measurement noise increases the corresponding proportion. The effect of the variance of the process noise on $\hat x_{k|k}^i$ is similar to the analysis above. The DKF algorithm adjusts the corresponding weights according to process noise and measurement noise in a similar way to that of the KF algorithm.

Using a similar approach, the state update method of MCKF algorithm \cite{chen2017maximum} can be expressed as
\begin{align}
\left\{ {\begin{array}{*{20}{l}}
  \begin{gathered}
  {\boldsymbol{\hat x}}_{k|k}^i = {{{\boldsymbol{\bar A}}}_{\boldsymbol{P}}}{\boldsymbol{\hat x}}_{k|k - 1}^i + {{{\boldsymbol{\bar A}}}_{{\boldsymbol{\bar R}}}}{{\boldsymbol{y}}_k}, \hfill \\
  {{{\boldsymbol{\bar A}}}_{\boldsymbol{P}}} = {\left[ {{{\left( {{\boldsymbol{P}}_{k|k - 1}^i} \right)}^{ - 1}} + {{\boldsymbol{C}}^T}{\boldsymbol{\bar R}}_k^{ - 1}{\boldsymbol{C}}} \right]^{ - 1}}{\left( {{\boldsymbol{P}}_{k|k - 1}^i} \right)^{ - 1}}, \hfill \\ 
\end{gathered}  \\ 
  {{{{\boldsymbol{\bar A}}}_{\boldsymbol{R}}} = {{\left[ {{{\left( {{\boldsymbol{P}}_{k|k - 1}^i} \right)}^{ - 1}} + {{\boldsymbol{C}}^T}{\boldsymbol{\bar R}}_k^{ - 1}{\boldsymbol{C}}} \right]}^{ - 1}}{{\boldsymbol{C}}^T}{\boldsymbol{\bar R}}_k^{ - 1},} 
\end{array}} \right.
\end{align}
where ${{\boldsymbol{\bar R}}_k} = {{\boldsymbol{B}}_{r;k}}{\boldsymbol{C}}_{y;k}^{ - 1}{\boldsymbol{B}}_{r;k}^T$ and ${{\boldsymbol{C}}_{y;k}} = diag\left[ {{G_\sigma }\left( {{e_{n + 1;k}}} \right), \cdots {G_\sigma }\left( {{e_{n + m;k}}} \right)} \right]$, and ${{\boldsymbol{B}}_{r;k}}$ can be obtained using the Cholesky decomposition of the variance ${{\boldsymbol{R}}_{I;k}}$ with impulse noise. ${{\boldsymbol{R}}_{I;k}}$ denotes the variance of the sequence that contains the impulse noise. It is easy to get
\begin{align}\label{rikjjrnkjj}
{[{{\boldsymbol{R}}_{I;k}}]_{jj}} > {[{{\boldsymbol{R}}_{N;k}}]_{jj}},
\end{align}
where ${{\boldsymbol{R}}_{N;k}}$ denotes the variance of the sequence that does not contain the impulse noise, and ${[ \cdot ]_{jj}}$ denotes the $j$th row and $j$th column element of a matrix. For the case in which both the state vector and the observation vector are one-dimensional, and when the error is affected by impulse noise, the weight adjustment factor ${C_{y;k}}$ is much less than 1, which significantly reduces the weight of ${y_k}$ in updated state $\hat x_{k|k}^i$, and the effect of impulse noise is thus mitigated. This adjustment mechanism is similar to that of conventional KF when the variance of the observed noise is large. When the noise is non-impulse noise, the weight adjustment factor $C_{y;k}^{ - 1} = 1$ attempts to raise the weight ${{\bar A}_{\bar R}}$ of the measurement information in $\hat x_{k|k}^i$ by the Gaussian Gaussian kernel function. However, since the kernel width cannot be set to infinity to account for the performance of the algorithm in non-Gaussian noise conditions, $C_{y;k}^{ - 1} > 1$ always holds. Furthermore, the non-Gaussian character of the measurement noise allows equation \eqref{rikjjrnkjj} to hold. 

The main reason for this situation is the limited downward adjustment of the Gaussian kernel function with a fixed kernelwidth. In other words, the covariance, which reflects the second-order statistical properties of non-Gaussian noise, is not sufficiently flexible to adjust the weights in the case of non-Gaussian noise, and which also provides an idea for improvement of the MCKF algorithm, i.e. using ${{{\boldsymbol{R}}_{N;k}}}$ instead of ${{\boldsymbol{R}}_{I;k}}$ to improve the flexibility of the weight adjustment of the algorithm. The detailed implementation of the proposed algorithms is shown in Section \ref{firstsupart}.

\begin{remark}
To reduce the negative impact of inappropriate covariance ${{\boldsymbol{R}}_{N;k}}$, it is also beneficial to increase the value of the kernelwidth appropriately so that the value of $C_{y;k}^{ - 1}$ is close to 1 when the errors are not caused by impulse noise, thus giving a more reasonable weight to ${{\bar A}_{\bar R}}$. This strategy also implies that the choice of kernelwidth for the M-MCKF algorithm is greater than that for the MCKF algorithm.
\end{remark}

\subsection{Algorithm derivation}\label{firstsupart}
First, the measurement noise with outliers is decomposed into Gaussian components with different means, variances, and proportions. The Gaussian mixture model (GMM) \cite{bishop2006pattern}, in this paper, is employed to decompose the measurement noise. The outlier-contaminated measurement noise is decomposed into
\begin{align}\label{RifokdmuifoK}
p\left( {{\boldsymbol{v}}_k^i} \right) = \sum\limits_{o = 1}^O {{\beta ^{i;o}}g\left( {{\boldsymbol{v}}_k^i|{\boldsymbol{\mu }}_k^{i;o},{\boldsymbol{R}}_k^{i;o}} \right)} 
\end{align}
with
\begin{align}\label{RiMUIOFjVikRioK}
\begin{gathered}
  g\left( {{\boldsymbol{v}}_k^i|{\boldsymbol{\mu }}_k^{i;o},{\boldsymbol{R}}_k^{i;o}} \right) =  \hfill \\
  \frac{{\exp \left\{ { - \frac{1}{2}{{\left[ {{\boldsymbol{v}}_k^i - {\boldsymbol{\mu }}_k^{i;o}} \right]}^T}{{\left( {{\boldsymbol{R}}_k^{i;o}} \right)}^{ - 1}}\left[ {{\boldsymbol{v}}_k^i - {\boldsymbol{\mu }}_k^{i;o}} \right]} \right\}}}{{{{\left( {2\pi } \right)}^{{m_i}/2}}{{\left| {{\boldsymbol{R}}_k^{i;o}} \right|}^{1/2}}}}, \hfill \\ 
\end{gathered}  
\end{align}
where $p\left( {{\boldsymbol{v}}_k^i} \right)$ is the probability density function (PDF) of the measurement noise of the $i$th node, $g({\boldsymbol{v}}_k^i|{\boldsymbol{\mu }}_k^{i;o},{\boldsymbol{R}}_k^{i;o})$ denotes the Gaussian distribution with mean ${{\boldsymbol{\mu }}_k^{i;o}}$ and variance ${{\boldsymbol{R}}_k^{i;o}}$, and $O$ is the number of the Gaussian component; ${{\beta ^{i;o}}}$ is the proportion of the $o$th component and $\left|  \cdot  \right|$ denote the determinant of the matrix. By using GMM, outliers can be represented as a Gaussian distribution with a large variance ${\boldsymbol{R}}_k^{i;l}$, and another Gaussian component with a relatively small variance ${\boldsymbol{R}}_k^{i;s}$. The variance ${\boldsymbol{R}}_k^{i;s}$ is employed to replace the variances in the MCKF and DMCKF algorithms, and one can obtain
\begin{align}
{\boldsymbol{R}}_k^{{\mho _i}} = diag\left[ {{\boldsymbol{R}}_k^{1;s},{\boldsymbol{R}}_k^{2;s}, \cdots {\boldsymbol{R}}_k^{{m_i};s}} \right].
\end{align}

The measurement equation and filter update will be reformulated as a regression problem in the linear-regression-based KF solution \cite{karlgaard2015nonlinear}. Denote ${{\boldsymbol{x}}_k}$ the real state of the target, and hence the state prediction error can be written as ${\boldsymbol{\varepsilon }}_{k|k - 1}^i = {{\boldsymbol{x}}_k} - {\boldsymbol{\hat x}}_{k|k - 1}^i$. With ${\boldsymbol{s}}_k^{{\mho _i}}$ and ${\boldsymbol{\hat x}}_{k|k - 1}^i$, the regression problem with the form of
\begin{align}\label{equ:kigjkixkio}
\left[ {\begin{array}{*{20}{c}}
  {{\boldsymbol{\hat x}}_{k|k - 1}^i} \\ 
  {{\boldsymbol{s}}_k^{{\mho _i}}} 
\end{array}} \right] = \left[ {\begin{array}{*{20}{c}}
  {{{\boldsymbol{I}}_n}} \\ 
  {{\boldsymbol{D}}_{\gamma ;k}^{{\mho _i}}{{\boldsymbol{C}}^{{\mho _i}}}} 
\end{array}} \right]{\boldsymbol{x}}_k^i{\text{ + }}{\boldsymbol{g}}_k^i,
\end{align}
where ${{{{\boldsymbol{I}}_n}}}$ represents an n-dimensional identity matrix and ${{\boldsymbol{g}}_k^i}$ represents the augmented noise vector containing the state and measurement errors of the dynamical system, which is defined as 
\begin{equation}
\begin{split}
{\boldsymbol{g}}_k^i = \left[ {\begin{array}{*{20}{c}}
  { - {\boldsymbol{\varepsilon }}_{k|k - 1}^i} \\ 
  {{\boldsymbol{D}}_{\gamma ;k}^{{\mho _i}}{\boldsymbol{v}}_k^{{\mho _i}}} 
\end{array}} \right].
\end{split}
\end{equation}
Assuming that the covariance matrix of the augmented vector ${{\text{E}}[{\boldsymbol{g}}_k^i{({\boldsymbol{g}}_k^i)^{\text{T}}}]}$ is positive definite yields the following:
\begin{equation}\label{equ:tkikbkikb}
\begin{split}
\begin{gathered}
  {\text{E}}\left[ {{\boldsymbol{g}}_k^i{{\left( {{\boldsymbol{g}}_k^i} \right)}^{\text{T}}}} \right] = \left[ {\begin{array}{*{20}{c}}
  {{\boldsymbol{P}}_{k|k - 1}^i}&0 \\ 
  0&{{\boldsymbol{D}}_{p;k}^{{\mho _i}}{\boldsymbol{R}}_k^{{\mho _i}}{\boldsymbol{D}}_{p;k}^{{\mho _i}}} 
\end{array}} \right] \hfill \\
   = \left[ {\begin{array}{*{20}{c}}
  {{\boldsymbol{B}}_{P\left( {k|k - 1} \right)}^i{{\left( {{\boldsymbol{B}}_{P\left( {k|k - 1} \right)}^i} \right)}^{\text{T}}}}&0 \\ 
  0&{{\boldsymbol{B}}_{R;k}^i{{\left( {{\boldsymbol{B}}_{R;k}^i} \right)}^{\text{T}}}} 
\end{array}} \right] \hfill \\
   = {\boldsymbol{B}}_k^i{\left( {{\boldsymbol{B}}_k^i} \right)^{\text{T}}}, \hfill \\ 
\end{gathered} 
\end{split}
\end{equation}
with 
\begin{equation}
\begin{split}
{\text{E}}[{\boldsymbol{D}}_{\gamma ;k}^{{\mho _i}}] = \operatorname{diag} {\{ p_k^{i;j}{{\boldsymbol{I}}_{{m_j}}}\} _{j \in {\mho _i}}} = {\boldsymbol{D}}_{p;k}^{{\mho _i}},
\end{split}
\end{equation}
where, ${{\boldsymbol{B}}_k^i}$ and ${{({\boldsymbol{B}}_k^i)^{\text{T}}}}$ can be produced by the Cholesky decomposition of ${{\text{E}}[{\boldsymbol{g}}_k^i{({\boldsymbol{g}}_k^i)^{\text{T}}}]}$, and matrix ${{\boldsymbol{D}}_{p;k}^{{\mho _i}}}$ represents the expectation of ${{{\boldsymbol{D}}_{\gamma ;k}^{{\mho _i}}}}$. Left multiplying each term of (\ref{equ:kigjkixkio}) by $ {{({\boldsymbol{B}}_k^i)^{ - 1}}}$ yields
\begin{equation}
\begin{split}\label{dwxeikikik3}
{\boldsymbol{d}}_k^i = {\boldsymbol{W}}_k^i{\boldsymbol{x}}_k^i + {\boldsymbol{e}}_k^i,
\end{split}
\end{equation}
where 
\begin{equation}\label{equ:jkiociuf}
\begin{split}
\left\{ {\begin{array}{*{20}{l}}
  {{\boldsymbol{d}}_k^i = {{\left( {{\boldsymbol{B}}_k^i} \right)}^{ - 1}}\left[ {\begin{array}{*{20}{c}}
  {{\boldsymbol{\hat x}}_{k|k - 1}^i} \\ 
  {{\boldsymbol{s}}_k^{{\mho _i}}} 
\end{array}} \right],} \\ 
  {{\boldsymbol{W}}_k^i = {{\left( {{\boldsymbol{B}}_k^i} \right)}^{ - 1}}\left[ {\begin{array}{*{20}{c}}
  {\boldsymbol{I}} \\ 
  {{\boldsymbol{D}}_{\gamma ;k}^{{\mho _i}}{{\boldsymbol{C}}^{{\mho _i}}}} 
\end{array}} \right],} \\ 
  {{\boldsymbol{e}}_k^i = {{\left( {{\boldsymbol{B}}_k^i} \right)}^{ - 1}}{\boldsymbol{g}}_k^i.} 
\end{array}} \right.
\end{split}
\end{equation}

The following cost function based on the MC criterion is suggested by the aforementioned derivation:
\begin{equation}
\begin{split}
{J}\left( {x_k^i} \right) = \frac{1}{H}\sum\limits_{h = 1}^H {{\operatorname{G} _\sigma }\left( {d_k^{i;h} - {\boldsymbol{w}}_k^{i;h}{\boldsymbol{x}}_k^i} \right)},
\end{split}
\end{equation}
where ${d_k^{i;h}}$ represents the ${h}$th element of ${{\boldsymbol{d}}_k^i}$, ${{\boldsymbol{w}}_k^{i;h}}$ represents the ${h}$th row  of ${{\boldsymbol{W}}_k^i}$, and $H = n + {m_{i;a}}$ (${\left( {{m_{i;a}}} \right)_{i \in \Omega }} = \sum\nolimits_{j \in {\mho _i}} {{m_j}} $) denotes the number of elements of ${{\boldsymbol{d}}_k^i}$. Then, the objective function of the optimal state estimation ${{\boldsymbol{x}}_k^i}$ based on the MC criterion is
\begin{equation}
\begin{split}
\hat x_k^i = \mathop {\max }\limits_{x_k^i} J\left( {x_k^i} \right) = \mathop {\max }\limits_{x_k^i} \sum\limits_{h = 1}^L {{G_\sigma }} \left( {e_k^{i;h}} \right),
\end{split}
\end{equation}
with
\begin{align}\label{equ:tkkikihw}
e_k^{i;h} = d_k^{i;h} - {\boldsymbol{w}}_k^{i;h}{\boldsymbol{x}}_k^i,
\end{align}
where  ${{h = 1,2, \cdots H}}$ is the ${h}$th element of ${e_k^{i}}$. Finally, the optimal state estimation of ${{\boldsymbol{\hat x}}_k^i}$ can be achieved by maximizing the information potential $V\left(  \cdot  \right)$. To that end, we set the gradient of the cost function ${{{J}({\boldsymbol{x}}_k^i)}}$ regarding ${{{\boldsymbol{x}}_k^i}}$  to zero,
and obtaining the optimal state of ${{{\boldsymbol{x}}_k^i}}$ is relatively easy:
\begin{equation}\label{equ:khfidtkkhf}
\begin{split}
\begin{gathered}
  {\boldsymbol{x}}_k^i = {\left[ {\sum\limits_{h = 1}^H {{G_\sigma }\left( {e_k^{i;h}} \right)} {{\left( {{\boldsymbol{w}}_k^{i;h}} \right)}^T}{\boldsymbol{w}}_k^{i;h}} \right]^{ - 1}} \hfill \\
   \times \sum\limits_{h = 1}^H {{G_\sigma }\left( {e_k^{i;h}} \right)} {\left( {{\boldsymbol{w}}_k^{i;h}} \right)^T}d_k^{i;h}. \hfill \\ 
\end{gathered} 
\end{split}
\end{equation}
Since ${e_k^{i;h} = d_k^{i;h} - {\boldsymbol{w}}_k^{i;h}{\boldsymbol{x}}_k^i}$ is a function of ${{\boldsymbol{x}}_k^i}$, the optimal state estimation in (\ref{equ:khfidtkkhf}) is a fixed-point iterative equation of ${{\boldsymbol{x}}_k^i}$, and can be expressed in the following form:
\begin{equation}\label{equ:dkixkkfddkix}
\begin{split}
{\boldsymbol{x}}_k^i = {\boldsymbol{f}}\left( {{\boldsymbol{x}}_k^i} \right),
\end{split}
\end{equation}
where
\begin{equation}
\begin{split}
\begin{gathered}
  {\boldsymbol{f}}\left( {{\boldsymbol{x}}_k^i} \right) = {\left[ {\sum\limits_{h = 1}^H {{G_\sigma }\left( {d_k^{i;h} - {\boldsymbol{w}}_k^{i;h}{\boldsymbol{x}}_k^i} \right)} {{\left( {{\boldsymbol{w}}_k^{i;h}} \right)}^{\text{T}}}{\boldsymbol{w}}_k^{i;h}} \right]^{ - 1}} \hfill \\
   \times \sum\limits_{h = 1}^H {{G_\sigma }\left( {d_k^{i;h} - {\boldsymbol{w}}_k^{i;h}{\boldsymbol{x}}_k^i} \right)} {\left( {{\boldsymbol{w}}_k^{i;h}} \right)^{\text{T}}}d_k^{i;h}. \hfill \\ 
\end{gathered} 
\end{split}
\end{equation}

According to the above derivation, the iterative equation in (\ref{equ:dkixkkfddkix}) can be written as follows:
\begin{equation}
\begin{split}
{\left( {{\boldsymbol{\hat x}}_k^i} \right)_{t + 1}} = {\boldsymbol{f}}\left[ {{{\left( {{\boldsymbol{x}}_k^i} \right)}_t}} \right].
\end{split}
\end{equation}
Here, ${{({\boldsymbol{\hat x}}_k^i)_{t + 1}}}$ represents the result of ${{\boldsymbol{x}}_k^i}$ at the fixed-point iteration ${t+1}$, and (\ref{equ:khfidtkkhf}) can be further written in the form of matrix multiplication:
\begin{equation}\label{equ:kidkiatkwi}
\begin{split}
{\left( {{\boldsymbol{x}}_k^i} \right)_{t + 1}} = {\left[ {{{\left( {{\boldsymbol{W}}_k^i} \right)}^{\text{T}}}{{\left( {{\boldsymbol{\Lambda }}_k^i} \right)}_t}{\boldsymbol{W}}_k^i} \right]^{ - 1}}{\left( {{\boldsymbol{W}}_k^i} \right)^{\text{T}}}{\left( {{\boldsymbol{\Lambda }}_k^i} \right)_t}{\boldsymbol{d}}_k^i,
\end{split}
\end{equation}
where 
\begin{subequations}\label{equ:jkktjsikesg}
\begin{numcases}{}
{\left( {{\boldsymbol{\Lambda }}_k^i} \right)_t} = \left[ {\begin{array}{*{20}{c}}
  {{{\left( {{\boldsymbol{\Lambda }}_{x;k}^i} \right)}_t}}&0 \\ 
  0&{{{\left( {{\boldsymbol{\Lambda }}_{y;k}^i} \right)}_t}} 
\end{array}} \right], \\ 
{({\boldsymbol{\Lambda }}_{x;k}^i)_t} = {\text{diag}}\left[ {{G_\sigma }[{{(e_k^{i;1})}_t}], \cdots ,{G_\sigma }[{{(e_k^{i;n})}_t}]} \right],\\  \label{equ:kinegkie}
{({\boldsymbol{\Lambda }}_{y;k}^i)_t} = {\text{diag}}\left[ {{G_\sigma }[{{(e_k^{i;n + 1})}_t}], \cdots ,{G_\sigma }[{{(e_k^{i;n + {m_{i;a}}})}_t}]} \right].  \label{equ:kngegknig}
\end{numcases}
\end{subequations}
According to (\ref{equ:jkiociuf}), (\ref{equ:kidkiatkwi}), and (\ref{equ:jkktjsikesg}), we obtain the following:
\begin{align}
\begin{gathered}
  {[{({\boldsymbol{W}}_k^i)^{\text{T}}}{({\boldsymbol{\Lambda }}_k^i)_t}{\boldsymbol{W}}_k^i]^{ - 1}} =  \hfill \\
  {\left\{ \begin{gathered}
  {[{({\boldsymbol{B}}_{P\left( {k|k - 1} \right)}^i)^{ - 1}}]^{\text{T}}}{({\boldsymbol{\Lambda }}_{x;k}^i)_t}{({\boldsymbol{B}}_{P\left( {k|k - 1} \right)}^i)^{ - 1}} +  \hfill \\
  {({\boldsymbol{D}}_{\gamma ;k}^{{\mho _i}}{{\boldsymbol{C}}^{{\mho _i}}})^{\text{T}}}{[{({\boldsymbol{B}}_{R;k}^i)^{ - 1}}]^{\text{T}}}{({\boldsymbol{\Lambda }}_{y;k}^i)_t}{({\boldsymbol{B}}_{R;k}^i)^{ - 1}}({\boldsymbol{D}}_{\gamma ;k}^{{\mho _i}}{{\boldsymbol{C}}^{{\mho _i}}}) \hfill \\ 
\end{gathered}  \right\}^{ - 1}} .\hfill \\ 
\end{gathered} 
\end{align}
We then apply the following matrix inversion lemma \cite{chen2017maximum,HE20221362}, then define
\begin{equation}
\begin{split}
\left\{ \begin{gathered}
  {\boldsymbol{G}} = {\left[ {{{\left( {{\boldsymbol{B}}_{P\left( {k|k - 1} \right)}^i} \right)}^{ - 1}}} \right]^{\text{T}}}{\left( {{\boldsymbol{\Lambda }}_{x;k}^i} \right)_t}{\left( {{\boldsymbol{B}}_{P\left( {k|k - 1} \right)}^i} \right)^{ - 1}}, \hfill \\
  {\boldsymbol{B}} = {\left( {{\boldsymbol{D}}_{\gamma ;k}^{{\mho _i}}{{\boldsymbol{C}}^{{\mho _i}}}} \right)^{\text{T}}}, \hfill \\
  {\boldsymbol{C}} = {\left[ {{{\left( {{\boldsymbol{B}}_{R;k}^i} \right)}^{ - 1}}} \right]^{\text{T}}}{\left( {{\boldsymbol{\Lambda }}_{y;k}^i} \right)_t}{\left( {{\boldsymbol{B}}_{R;k}^i} \right)^{ - 1}}, \hfill \\
  {\boldsymbol{D}} = {\boldsymbol{D}}_{\gamma ;k}^{{\mho _i}}{{\boldsymbol{C}}^{{\mho _i}}}, \hfill \\ 
\end{gathered}  \right.
\end{split}
\end{equation}
and obtain the expression (\ref{equ:jt1jkpkib}) given below.
\newcounter{mytempeqncnt}
\begin{figure*}[!t]
\normalsize
\setcounter{mytempeqncnt}{\value{equation}}
\begin{equation}\label{equ:jt1jkpkib}
\begin{gathered}
\begin{gathered}
  {\left[ {{{\left( {{\boldsymbol{W}}_k^i} \right)}^{\text{T}}}{{\left( {{\boldsymbol{\Lambda }}_k^i} \right)}_t}{\boldsymbol{W}}_k^i} \right]^{ - 1}} = {\boldsymbol{B}}_{P\left( {k|k - 1} \right)}^i\left( {{\boldsymbol{\Lambda }}_{x;k}^i} \right)_t^{ - 1}{\left( {{\boldsymbol{B}}_{P\left( {k|k - 1} \right)}^i} \right)^T} - {\boldsymbol{B}}_{P\left( {k|k - 1} \right)}^i\left( {{\boldsymbol{\Lambda }}_{x;k}^i} \right)_t^{ - 1}{\left( {{\boldsymbol{B}}_{P\left( {k|k - 1} \right)}^i} \right)^T}{\left( {{\boldsymbol{D}}_{\gamma ;k}^{{\mho _i}}{C^{{\mho _i}}}} \right)^T} \hfill \\
  \left[ {{\boldsymbol{B}}_{R;k}^i\left( {{\boldsymbol{\Lambda }}_{y;k}^i} \right)_t^{ - 1}{{\left( {{\boldsymbol{B}}_{R;k}^i} \right)}^T} + \left( {{\boldsymbol{D}}_{\gamma ;k}^{{\mho _i}}{{\boldsymbol{C}}^{{\mho _i}}}} \right){\boldsymbol{B}}_{P\left( {k|k - 1} \right)}^i\left( {{\boldsymbol{\Lambda }}_{x;k}^i} \right)_t^{ - 1}{{\left( {{\boldsymbol{B}}_{P\left( {k|k - 1} \right)}^i} \right)}^T}{{\left( {{\boldsymbol{D}}_{\gamma ;k}^{{\mho _i}}{{\boldsymbol{C}}^{{\mho _i}}}} \right)}^T}} \right] \times  \hfill \\
  \left( {{\boldsymbol{D}}_{\gamma ;k}^{{\mho _i}}{{\boldsymbol{C}}^{{\mho _i}}}} \right){\boldsymbol{B}}_{P\left( {k|k - 1} \right)}^i\left( {{\boldsymbol{\Lambda }}_{x;k}^i} \right)_t^{ - 1}{\left( {{\boldsymbol{B}}_{P\left( {k|k - 1} \right)}^i} \right)^T} \hfill \\ 
\end{gathered}     
\end{gathered} 
\end{equation}
\hrulefill
\vspace*{4pt}
\end{figure*}
According to (\ref{equ:jkiociuf}), (\ref{equ:kidkiatkwi}), and (\ref{equ:jkktjsikesg}), we obtain the following:
\begin{equation}\label{equ:kious1fkrikb}
\begin{split}
\begin{gathered}
  {({\boldsymbol{W}}_k^i)^{\text{T}}}{({\boldsymbol{\Lambda }}_k^i)_t}{\boldsymbol{d}}_k^i = {[{({\boldsymbol{B}}_{P\left( {k|k - 1} \right)}^i)^{ - 1}}]^T}{({\boldsymbol{\Lambda }}_{x;k}^i)_t}{({\boldsymbol{B}}_{P\left( {k|k - 1} \right)}^i)^{ - 1}} \hfill \\
   \times {\boldsymbol{\hat x}}_{k|k - 1}^i + {({\boldsymbol{D}}_{\gamma ;k}^{{\mho _i}}{{\boldsymbol{C}}^{{\mho _i}}})^T}{[{({\boldsymbol{B}}_{R;k}^i)^{ - 1}}]^T}{({\boldsymbol{\Lambda }}_{y;k}^i)_t}{({\boldsymbol{B}}_{R;k}^i)^{ - 1}}{\boldsymbol{s}}_k^{{\mho _i}}. \hfill \\ 
\end{gathered} 
\end{split}
\end{equation}
Substituting formulas (\ref{equ:jt1jkpkib}) and (\ref{equ:kious1fkrikb}) into (\ref{equ:kidkiatkwi}) yields
\begin{equation}\label{equ:tijksxiug}
\begin{split}
{({\boldsymbol{\hat x}}_{k|k}^i)_{t + 1}} = {\boldsymbol{\hat x}}_{k|k - 1}^i + {({\boldsymbol{\bar K}}_k^i)_t}({\boldsymbol{s}}_k^{{\mho _i}} - {\boldsymbol{D}}_{\gamma ;k}^{{\mho _i}}{{\boldsymbol{C}}^{{\mho _i}}}{\boldsymbol{\hat x}}_{k|k - 1}^i),
\end{split}
\end{equation}
where 
\begin{subequations}\label{463KPRiiikkkT}
\begin{numcases}{}
\begin{gathered}\label{equ:kirptikc}
  {({\boldsymbol{\bar K}}_k^i)_t} = {({\boldsymbol{\bar P}}_{k|k - 1}^i)_t}{({\boldsymbol{D}}_{\gamma ;k}^{{\mho _i}}{{\boldsymbol{C}}^{{\mho _i}}})^T} \times  \hfill \\
  {\left[ {{\boldsymbol{D}}_{\gamma ;k}^{{\mho _i}}{{\boldsymbol{C}}^{{\mho _i}}}{{({\boldsymbol{\bar P}}_{k|k - 1}^i)}_t}{{({\boldsymbol{D}}_{\gamma ;k}^{{\mho _i}}{{\boldsymbol{C}}^{{\mho _i}}})}^T} + {{({\boldsymbol{\bar R}}_k^i)}_t}} \right]^{ - 1}}, \hfill \\ 
\end{gathered} \\ 
{({\boldsymbol{\bar P}}_{k|k - 1}^i)_t} = {\boldsymbol{B}}_{P\left( {k|k - 1} \right)}^i({\boldsymbol{\Lambda }}_{x;k}^i)_t^{ - 1}{({\boldsymbol{B}}_{P\left( {k|k - 1} \right)}^i)^{\text{T}}},\\  \label{equ:trkibyi}
{({\boldsymbol{\bar R}}_k^i)_t} = {\boldsymbol{B}}_{R;k}^i({\boldsymbol{\Lambda }}_{y;k}^i)_t^{ - 1}{({\boldsymbol{B}}_{R;k}^i)^{\text{T}}}.  \label{equ:tkrikyi}
\end{numcases}
\end{subequations}
The following updates are made iteratively to the posterior covariance:
\begin{align}\label{pkkupidateIJ}
\begin{gathered}
  {\boldsymbol{P}}_{k|k}^i = ({\boldsymbol{I}} - {\boldsymbol{\tilde K}}_k^i{\boldsymbol{D}}_{\gamma ;k}^{{\mho _i}}{{\boldsymbol{C}}^{{\mho _i}}}){\boldsymbol{P}}_{k|k - 1}^i{({\boldsymbol{I}} - {\boldsymbol{\tilde K}}_k^i{\boldsymbol{D}}_{\gamma ;k}^{{\mho _i}}{{\boldsymbol{C}}^{{\mho _i}}})^T} \hfill \\
   + {\boldsymbol{\tilde K}}_k^i{\boldsymbol{R}}_k^{{\mho _i}}{({\boldsymbol{\tilde K}}_k^i)^T}. \hfill \\ 
\end{gathered} 
\end{align}
\begin{remark}
Equation (\ref{equ:tijksxiug}) is the optimal solution of ${{\boldsymbol{x}}_k^i}$, and it depends on the prior estimate ${{\boldsymbol{\hat x}}_{k|k - 1}^i}$ and available observation information ${{{\boldsymbol{s}}_k^{{\mho _i}}}}$. The observation information obtainable by the ${i}$th node is determined by the matrix ${{{\boldsymbol{D}}_{\gamma ;k}^{{\mho _i}}}}$.
\end{remark}
\begin{remark}
If the noise is not decomposed and the obtained ${{\boldsymbol{R}}_{I;k}}$ is used directly, then the algorithm proposed in this paper will degenerate into the D-MCKF-DPD algorithm.
\end{remark}

According to the above derivations, we summarize the steps of the proposed DM-MCKF-DPD algorithm as follows.
\begin{algorithm}[t]
\caption{DM-MCKF-DPD} \label{DMCKF-DPD}
\LinesNumbered 
\KwIn{state transition matrix ${{\boldsymbol{A}}_k}$, measurement matrix ${{\boldsymbol{C}}^i}$, covariance matrices ${{\boldsymbol{Q}}_{k - 1}}$, ${\boldsymbol{R}}_k^i$, and kernelwidth $\sigma$}
\KwOut{${\boldsymbol{\hat x}}_{k|k}^i$}
\textbf{Initialization}: Initialize the values of the ${\sigma }$, ${\varepsilon }$ (a small positive number), ${{\boldsymbol{\hat x}}_{0|0}^i}$, and ${{\boldsymbol{P}}_{0|0}^i}$\; 
\For{$k\leftarrow $1$ $ \KwTo $K$}{ 
Calculate ${\boldsymbol{\hat x}}_{k|k - 1}^i$ and ${\boldsymbol{P}}_{k|k - 1}^i$ using \eqref{akkj1dAkJJ} and \eqref{akk1234}\;
Calculate ${\boldsymbol{R}}_k^{{\mho _i}}$ using \eqref{RifokdmuifoK} and \eqref{RiMUIOFjVikRioK}\;
Calculate ${{\boldsymbol{B}}_{P\left( {k|k - 1} \right)}^i}$ and ${\boldsymbol{B}}_{R;k}^i$ using Cholesky decomposition\;
Let $t=1$ and ${({\boldsymbol{\hat x}}_{k|k}^i)_0} = {\boldsymbol{\hat x}}_{k|k - 1}^i$ to initialize the fixed-point iteration strategy\;
\While{$||{({\boldsymbol{\hat x}}_{k|k}^i)_{t + 1}} - {({\boldsymbol{\hat x}}_{k|k}^i)_t}||/{({\boldsymbol{\hat x}}_{k|k}^i)_t} \geqslant \varepsilon $}{
 update state ${({\boldsymbol{\hat x}}_{k|k}^i)_{t + 1}}$ using \eqref{equ:tijksxiug}, \eqref{463KPRiiikkkT}, \eqref{equ:jkktjsikesg}, \eqref{equ:jkiociuf}, and \eqref{equ:tkkikihw}\;
}
Update covariance matrix using \eqref{pkkupidateIJ}\;
}
\end{algorithm}

\subsection{Distributed consensus filter}
Based on the proposed DM-MCKF-DPD algorithm, the consensus filter algorithm is derived by fusing the updated state of the neighbor node at the previous time, and it is called the consensus-based DM-MCKF-DPD (C-DM-MCKF-DPD) algorithm. It is assumed that the measurement information ${\boldsymbol{y}}_k^j$ and updated status ${({\boldsymbol{\hat x}}_{k - 1|k - 1}^j)_{t + 1}}$ of neighbor node ${j \in {\Omega _i}}$ are transmitted to node $i$ through the same channel, thus, ${\gamma _{k - 1}^{i,j}}$ should be introduced to represent the case of packet loss in the updated state ${({\boldsymbol{\hat x}}_{k - 1|k - 1}^j)_{t + 1}}$. According to the above analysis, the consensus scheme can be expressed as
\begin{align}
\begin{gathered}
  {({\boldsymbol{\hat x}}_{k|k}^i)_{t + 1}} = {\boldsymbol{\hat x}}_{k|k - 1}^i + {({\boldsymbol{\bar K}}_k^i)_t}({\boldsymbol{s}}_k^{{\mho _i}} - {\boldsymbol{D}}_{\gamma ;k}^{{\mho _i}}{{\boldsymbol{C}}^{{\mho _i}}}{\boldsymbol{\hat x}}_{k|k - 1}^i) \hfill \\
   + \eta \sum\limits_{j \in {\Omega _i}} {\gamma _{k - 1}^{i,j}\left[ {{{({\boldsymbol{\hat x}}_{k|k}^j)}_{t + 1}} - {{({\boldsymbol{\hat x}}_{k|k}^i)}_{t + 1}}} \right]} . \hfill \\ 
\end{gathered} 
\end{align}
where $\eta  \geqslant 0$ is the consensus coefficient. The consensus scheme, which intuitively improves the estimated performance of node $i$ and reduces the difference between nodes, fuses the measurement information of the neighbour node through the Kalman gain ${({\boldsymbol{\bar K}}_k^i)_t}$ and fuses the estimated difference between the node $i$ and its neighbour node $j$ through the consensus coefficient. In particular, the consensus strategy dose not work when $\eta=0$.

\subsection{Computational complexity}
The computational complexity of the proposed DM-MCKF-DPD algorithm can be compared with that of the stationary DKF \cite{8409298} in terms of the equations and operations utilized by the two algorithms, which are presented in Table 1.

The stationary DKF algorithm mainly includes (\ref{equ:aa}), (\ref{equ:kiyikd}), \eqref{akkj1dAkJJ} and \eqref{akk1234} cited in the present paper, and (5) and (6) cited within\cite{8409298}. Therefore, we can evaluate the computational complexity of the stationary DKF as
\begin{equation}\label{equ:iugoiug}
\begin{split}
\begin{gathered}
  {S_{SDKF}} = 11{n^3} + 12m_{i;a}^2n + 10{m_{i;a}}{n^2} + 4m_{i;a}^2 \hfill \\
   - 2{m_{i;a}}n - {n^2} - 2n - {m_{i;a}} + 2O(m_{i;a}^3). \hfill \\ 
\end{gathered}  
\end{split}
\end{equation}
Accordingly, the computational complexity of the DM-MCKF-DPD algorithm can be defined based on an average number of fixed-point iterative algorithm iterations ${T}$ as 
\begin{equation}\label{equ:ijgotno}
\begin{split}
\begin{gathered}
  {S_{{\text{SDMCKF}}}}{\text{  =  6}}Tm_{i;a}^3 + 6T{n^3} + {\text{16}}Tm_{i;a}^2n +  \hfill \\
  10T{m_{i;a}}{n^2} + (2 - 3T)m_{i;a}^2 + 2{n^2} + (2 - 3T){m_{i;a}}n \hfill \\
   + ({\text{6}}T - 1){m_{i;a}} + ({\text{6}}T - 1)n + T{\text{O(}}{n^3}{\text{)}} + T{\text{O(}}m_{i;a}^3{\text{)}}. \hfill \\ 
\end{gathered} 
\end{split}
\end{equation}

\begin{table}\label{table1}
\caption{Computational complexities of the  proposed DM-MCKF-DPD algorithm and the stationary DKF algorithm\cite{8409298}}  
\begin{tabular*}{8.8cm}{lll}  
\hline  
Equation & Addition/subtraction     & Division,\\
         & and multiplication       & matrix inversion,\\ 
         &                          & Cholesky \\
         &                          & decomposition,\\
         &                          & and exponentiation\\
\hline  
(\ref{equ:aa}) & ${2n{m_{i;a}}}$                                 & 0\\
(\ref{equ:kiyikd}) & ${2m_{i;a}^2 - {m_{i;a}}}$            & 0\\
\eqref{akkj1dAkJJ} and \eqref{akk1234}   & ${2{n^2} - n}$                                  & 0\\
{(5)} in\cite{8409298}  & ${\begin{gathered}
  9{n^3} - 4{n^2} + 6{m_{i;a}}{n^2} +  \hfill \\
  4m_{i;a}^2n - 3{m_{i;a}}n + m_{i;a}^2 \hfill \\ 
\end{gathered} }$                                                                                      & ${{\text{O(}}m_{i;a}^3{\text{)}}}$\\
{(6a)} in\cite{8409298} & ${{\text{2}}m_{i;a}^2n + 3{m_{i;a}}n - n + 2{n^2}}$               & 0\\ 
{(6b)} in\cite{8409298} & ${\begin{gathered}
  {\text{2}}{n^3} - {n^2} + {\text{4}}{m_{i;a}}{n^2} +  \hfill \\
  {\text{6}}m_{i;a}^2n - {\text{4}}{m_{i;a}}n + m_{i;a}^2{\text{ }} \hfill \\ 
\end{gathered} }$                                                                                      & ${{\text{O(}}m_{i;a}^3{\text{)}}}$\\
(\ref{equ:epxeeg})   & ${2n + 4}$                                                                      & 2\\
(\ref{equ:tijksxiug})   & ${2m_{i;a}^2n + {\text{3}}{m_{i;a}}n}$                                          & 0\\
(\ref{equ:kirptikc})   & ${\begin{gathered}
  2m_{i;a}^3 + 8m_{i;a}^2n + 4{m_{i;a}}{n^2} -  \hfill \\
  5{m_{i;a}}n - m_{i;a}^2 \hfill \\ 
\end{gathered} }$                                                                       			  & 0\\
(\ref{equ:trkibyi})   & ${4{n^3} + {\text{4}}n}$                                        			  & ${2n + {\text{O}}({n^3}){\text{ }}}$\\
(\ref{equ:tkrikyi})   & ${2{m_{i;a}}n + 4{m_{i;a}} + 4m_{i;a}^3 - 2m_{i;a}^2}$           			  & ${2{m_{i;a}} + {\text{O}}(m_{i;a}^3)}$\\
(\ref{equ:kinegkie})   & ${2{n^2} + {\text{4}}n}$                                                      & ${2n}$\\  
(\ref{equ:kngegknig})   & ${2{m_{i;a}}n + {\text{4}}{m_{i;a}}}$                            		  & ${2{m_{i;a}}}$\\    
(\ref{equ:tkkikihw})   & ${2n{\text{ }}}$                                                 		       & 0\\   
\hline  
\end{tabular*}  
\end{table} 

We can infer from this discussion that the computational complexity of the DM-MCKF-DPD algorithm is moderate compared with that of the stationary DKF, provided that the value of ${T}$ is small, which is indeed the case, as will be demonstrated later in Section \ref{section:simulations}. 
\subsection{Convergence issue}
It is quite difficult to fully analyse the convergence behaviour of the DM- MCKF-DPD algorithm, which is based on the fixed-point iterative technique. Therefore, we merely provide a sufficient condition that guarantees the fixed-point iterative algorithm's convergence. However, a detailed proof process is not presented here because the convergence condition is similar to the analysis presented in an earlier work\cite{chen2017maximum}, which can be consulted for additional details.

\textbf{Theorem 1.} First, we assume the conditions ${\beta _i} > {\varsigma _i} = {{\sqrt n \sum\nolimits_{h = 1}^H {|d_k^{i;h}|||{{({\boldsymbol{w}}_k^{i;h})}^T}|{|_1}} } \mathord{\left/
 {\vphantom {{\sqrt n \sum\nolimits_{h = 1}^H {|d_k^{i;h}|||{{({\boldsymbol{w}}_k^{i;h})}^T}|{|_1}} } {{\chi _{\min }}\sum\nolimits_{h = 1}^H {{{({\boldsymbol{w}}_k^{i;h})}^T}{\boldsymbol{w}}_k^{i;h}} }}} \right.
 \kern-\nulldelimiterspace} {{\chi _{\min }}\sum\nolimits_{h = 1}^H {{{({\boldsymbol{w}}_k^{i;h})}^T}{\boldsymbol{w}}_k^{i;h}} }}$ and ${{\sigma _i} \ge \max {\rm{\{ }}\sigma _i^*,\sigma _i^\diamondsuit {\rm{\} }}}$. Here, ${\sigma _i^*}$ is the optimal result of the equation ${\phi ({\sigma _i}) = {\beta _i}}$, where 
\begin{align}\label{equ:kddokz}
\phi \left( {{\sigma _i}} \right) = \frac{{\sqrt n \sum\limits_{h = 1}^H {|d_k^{i;h}|||{{({\boldsymbol{w}}_k^{i;h})}^T}|{|_1}} }}{{{\chi _{\min }}\sum\limits_{h = 1}^H {{G_{{\sigma _i}}}({\beta _i}||{\boldsymbol{w}}_k^{i;h}|{|_1} + |d_k^{i;h}|)} {{({\boldsymbol{w}}_k^{i;h})}^T}{\boldsymbol{w}}_k^{i;h}}},
\end{align}, ${\sigma _i} \in \left( {0,\infty } \right)$, and 
\begin{align}\label{equ:wlindsisikc}
\psi \left( {{\sigma _i}} \right) = \frac{{\sqrt n \sum\limits_{h = 1}^H {\left[ \begin{gathered}
  ({\beta _i}||{\boldsymbol{w}}_k^{i;h}|{|_1} + |d_k^{i;h}|)||{\boldsymbol{w}}_k^{i;h}|{|_1} \times  \hfill \\
  ({\beta _i}||{({\boldsymbol{w}}_k^{i;h})^{\text{T}}}{\boldsymbol{w}}_k^{i;h}|{|_1} + ||{({\boldsymbol{w}}_k^{i;h})^{\text{T}}}d_k^{i;h}|{|_1}) \hfill \\ 
\end{gathered}  \right]} }}{{\sigma _i^2{\lambda _{\min }}\sum\limits_{h = 1}^H {{G_{{\sigma _i}}}({\beta _i}||{\boldsymbol{w}}_k^{i;h}|{|_1} + |d_k^{i;h}|){{({\boldsymbol{w}}_k^{i;h})}^{\text{T}}}{\boldsymbol{w}}_k^{i;h}} }}
\end{align}
and ${\sigma _i^\diamondsuit }$ is the result of the equation ${\psi ({\sigma _i}) = {\alpha _i}(0 < {\alpha _i} < 1)}$, where  ${\psi ({\sigma _i})}$ is given in (\ref{equ:wlindsisikc}) below. Accordingly, it holds that $||{\boldsymbol{f}}({\boldsymbol{x}}_k^i)|{|_1} \leqslant {\beta _i}$ and $||{\nabla _{{\boldsymbol{x}}_k^i}}{\boldsymbol{f}}({\boldsymbol{x}}_k^i)|{|_1} \leqslant {\alpha _i}$ for all ${{\boldsymbol{x}}_k^i \in \{ {\boldsymbol{x}}_k^i \in {\mathbb{R}^n}:||{\boldsymbol{x}}_k^i|{|_1} \leqslant {\beta _i}\} }$. Here, the ${n \times n}$ Jacobian matrix ${\boldsymbol{f}}({\boldsymbol{x}}_k^i)$ is given as follows:
\begin{equation}
\begin{split}
{\nabla _{{\boldsymbol{x}}_k^i}}{\boldsymbol{f}}\left( {{\boldsymbol{x}}_k^i} \right) = \left[ {\frac{\partial }{{\partial {\boldsymbol{x}}_k^{i;1}}}{\boldsymbol{f}}\left( {{\boldsymbol{x}}_k^i} \right) \cdots \frac{\partial }{{\partial {\boldsymbol{x}}_k^{i;n}}}{\boldsymbol{f}}\left( {{\boldsymbol{x}}_k^i} \right)} \right],
\end{split}
\end{equation} 
where the terms are defined in (\ref{equ:kddokzkkhiw}) below
\begin{align}\label{equ:kddokzkkhiw}
\begin{gathered}
  \frac{\partial }{{\partial {\boldsymbol{x}}_k^{i;j}}}{\boldsymbol{f}}\left( {{\boldsymbol{x}}_k^i} \right) = {{\boldsymbol{T}}_g}\frac{1}{{\sigma _i^2}}\sum\limits_{i = 1}^L {\left[ {e_k^{i;h}w_k^{i;h;j}{G_{{\sigma _i}}}\left( {e_k^i} \right){{\left( {{\boldsymbol{w}}_k^{i;h}} \right)}^{\text{T}}}d_k^{i;h}} \right]}  \hfill \\
   - {{\boldsymbol{T}}_g}\frac{1}{{\sigma _i^2}}\sum\limits_{i = 1}^L {\left[ {e_k^{i;h}w_k^{i;h;j}{G_{{\sigma _i}}}\left( {e_k^i} \right){{\left( {{\boldsymbol{w}}_k^{i;h}} \right)}^{\text{T}}}{\boldsymbol{w}}_k^{i;h}} \right]} {\boldsymbol{f}}\left( {{\boldsymbol{x}}_k^i} \right) \hfill \\ 
\end{gathered} 
\end{align}
with 
\begin{align}
{{\boldsymbol{T}}_g} = {\left[ {\sum\limits_{h = 1}^H {{\operatorname{G} _{{\sigma _i}}}\left( {d_k^{i;h} - {\boldsymbol{w}}_k^{i;h}{\boldsymbol{x}}_k^i} \right){{\left( {{\boldsymbol{w}}_k^{i;h}} \right)}^{\text{T}}}{\boldsymbol{w}}_k^{i;h}} } \right]^{ - 1}},
\end{align}
and ${{w_k^{i;h;j}}}$ is the ${j}$th element of the vector ${{{\boldsymbol{w}}_k^{i;h}}}$.

According to Theorem 1, we obtain the following conditions: 
\begin{align}
\left\{ {\begin{array}{*{20}{l}}
  {{{\left\| {{\boldsymbol{f}}\left( {{\boldsymbol{x}}_k^i} \right)} \right\|}_1} \leqslant {\beta _i},} \\ 
  {{{\left\| {{\nabla _{{\boldsymbol{x}}_k^i}}{\boldsymbol{f}}\left( {{\boldsymbol{x}}_k^i} \right)} \right\|}_1} \leqslant {\alpha _i} < 1,} 
\end{array}} \right.
\end{align}
if the kernelwidth ${\sigma }$ is sufficiently large (e.g., greater than ${\max {\text{\{ }}\sigma _i^*,\sigma _i^\diamondsuit {\text{\} }}}$). According to the Banach fixed-point theorem, the DM-MCKF-DPD fixed-point iterative technique will undoubtedly converge to a single fixed point in the range ${{\boldsymbol{x}}_k^i \in \{ {\boldsymbol{x}}_k^i \in {\mathbb{R}^n}: ||{\boldsymbol{x}}_k^i|{|_1} \leqslant {\beta _i}\} }$ provided that the initial state of the system meets the condition $||{({\boldsymbol{x}}_k^i)_0}|| \leqslant {\beta _i}$ and ${\sigma_i }$ is sufficiently large.

Theorem 1 demonstrates that the kernelwidth of the Gaussian kernel function has an important influence on the convergence of the DM-MCKF-DPD algorithm. Here, reducing the kernelwidth can improve the accuracy of state estimation, but this will also decrease the convergence rate of the algorithm or make it diverge. Conversely, increasing the kernel width will increase the convergence rate of the algorithm, but will often yield poor estimation performance under impulsive noise conditions. In practice, the kernelwidth can be selected by trial and error in accordance with the desired estimation accuracy and convergence rate of the algorithm.

\section{simulations}\label{section:simulations}
In this part, we compare the proposed algorithms with some existing algorithms, and the performance of these algorithms is adjusted to the optimum performance with the appropriate parameters. The performance of these algorithms is measured by the mean square deviation (MSD) in the form of
\begin{align}
MSD = 10{\log _{10}}\left\| {{\boldsymbol{x}}_k^i - {\boldsymbol{\hat x}}_k^i} \right\|.
\end{align}

Several noise models covered in this paper are presented before these simulations are implemented, such as Laplace noise, mixed-Gaussian noise, etc.
\begin{enumerate} 
    \item The mixed-Gaussian model \cite{HE2022,HE2023109188} takes the following form:
    \begin{align}
    v \sim \lambda \mathcal{N}\left( {{a_1},{\mu _1}} \right){\text{  +  }}\left( {1 - \lambda } \right)\mathcal{N}\left( {{a_2},{\mu _2}} \right),0 \leqslant \lambda  \leqslant 1,
    \end{align}
    where $\mathcal{N}\left( {{a_1},{\mu _1}} \right){\text{ }}$ denotes the Gaussian distribution with mean $a_1$ and variance ${{\mu _1}}$, and $\lambda $ represents the mixture coefficient of two kinds of Gaussian distribution. The mixed-Gaussian distribution can be abbreviated as $v \sim M\left( {\lambda ,{a_1},{a_2},{\mu _1},{\mu _2}} \right)$.
    \item The PDF of the Laplace distribution is $f\left( {v|\mu ,b} \right) = {1 \mathord{\left/
 {\vphantom {1 {2b}}} \right.
 \kern-\nulldelimiterspace} {2b}}\exp \left( { - {{\left| {v - \mu } \right|} \mathord{\left/
 {\vphantom {{\left| {v - \mu } \right|} b}} \right.
 \kern-\nulldelimiterspace} b}} \right)$ with a location parameter $\mu $ and a scale parameter $b$.
    \item The characteristic function of the $\alpha$-stable noise is presented in \cite{HE2022}, and the noise that obeys the $\alpha$-stable distribution is written as $v \sim S\left( {a,b,\gamma ,\varpi } \right)$, where parameter $a$, $b$, $\gamma$, and $\varpi $ are the characteristic factor, symmetry parameter, dispersion parameter, and location parameter.
\end{enumerate}

In this paper, four scenarios are considered, it is assumed that all of the scene's process noise is Gaussian noise with form ${\boldsymbol{q}} \sim \mathcal{N}\left( {0,0.01{{\boldsymbol{I}}_n}} \right)$. The distribution of the measurement noise for these four scenarios is $\mathcal{N}\left( {0,0.01{{\boldsymbol{I}}_{{m_i}}}} \right)$, $r \sim M\left( {0.9,0,0,0.01,64} \right)$, $f\left( {{\boldsymbol{q}}|0,9} \right)$, and $S\left( {1.2,1.0,0,0.5} \right)$, respectively.

A vehicle tracking model \cite{HE2022} is considered to evaluate the effectiveness of the M-MCKF and DM-MCKF algorithms, and the state space model is
\begin{align}
{\boldsymbol{x}}_k^i = \left[ {\begin{array}{*{20}{c}}
  1&0&{\Delta T}&0 \\ 
  0&1&0&{\Delta T} \\ 
  0&0&1&0 \\ 
  0&0&0&1 
\end{array}} \right]{\boldsymbol{x}}_{k - 1}^i + {{\boldsymbol{q}}_{k - 1}},
\end{align}
and
\begin{align}
{\boldsymbol{y}}_k^i = \left[ {\begin{array}{*{20}{c}}
  {\begin{array}{*{20}{c}}
  1&0&0&0 
\end{array}} \\ 
  {\begin{array}{*{20}{c}}
  \begin{gathered}
  0 \hfill \\
  0 \hfill \\
  0 \hfill \\ 
\end{gathered} &\begin{gathered}
  1 \hfill \\
  0 \hfill \\
  0 \hfill \\ 
\end{gathered} &\begin{gathered}
  0 \hfill \\
  1 \hfill \\
  0 \hfill \\ 
\end{gathered} &\begin{gathered}
  0 \hfill \\
  0 \hfill \\
  1 \hfill \\ 
\end{gathered}  
\end{array}} 
\end{array}} \right]{\boldsymbol{x}}_k^i + {\boldsymbol{v}}_k^i,
\end{align}
where ${\boldsymbol{x}}_k^i = {\left[ {\begin{array}{*{20}{c}}
  {x_{1;k}^i}&{x_{2;k}^i}&{x_{3;k}^i}&{x_{4;k}^i} 
\end{array}} \right]^T}$ is the state of the vehicle output by the $i$the node, ${x_{1;k}^i}$ and ${x_{2;k}^i}$ denote the position of vehicle on the $x$ and $y$ axis, ${x_{3;k}^i}$ and ${x_{4;k}^i}$ denote the velocity of vehicle on the $x$ and $y$ axis. $\Delta T = 0.1\sec $ is the time interval, and the covariance matrix of process noise is ${{\boldsymbol{Q}}_{k - 1}} = \left[ {\begin{array}{*{20}{c}}
  {\tfrac{{\Delta {T^2}}}{4}}&0&{\tfrac{{\Delta {T^3}}}{2}}&0 \\ 
  0&{\tfrac{{\Delta {T^2}}}{4}}&0&{\tfrac{{\Delta {T^3}}}{2}} \\ 
  {\tfrac{{\Delta {T^3}}}{2}}&0&{\Delta {T^2}}&0 \\ 
  0&{\tfrac{{\Delta {T^3}}}{2}}&0&{\Delta {T^2}} 
\end{array}} \right]$. The initial states of ${{\boldsymbol{x}}_0}$, ${{\boldsymbol{\hat x}}_{0|0}}$, and ${{\boldsymbol{\hat P}}_{0|0}}$ are 
\begin{align}\label{inistate}
\left\{ \begin{gathered}
  {{\boldsymbol{x}}_0} \sim \mathcal{N}\left( {0,{{\boldsymbol{I}}_n}} \right), \hfill \\
  {{{\boldsymbol{\hat x}}}_{0|0}} \sim \mathcal{N}\left( {{{\boldsymbol{x}}_0},{{\boldsymbol{I}}_n}} \right), \hfill \\
  {{\boldsymbol{P}}_{0|0}} = {{\boldsymbol{I}}_n}, \hfill \\ 
\end{gathered}  \right.
\end{align}
and the threshold is set to $\varepsilon  = {10^{ - 6}}$.

\subsection{Performance verification of the M-MCKF and DM-MCKF algorithms}
\begin{figure}[!t]
\centering
\subfloat[]{\includegraphics[width=0.5\columnwidth]{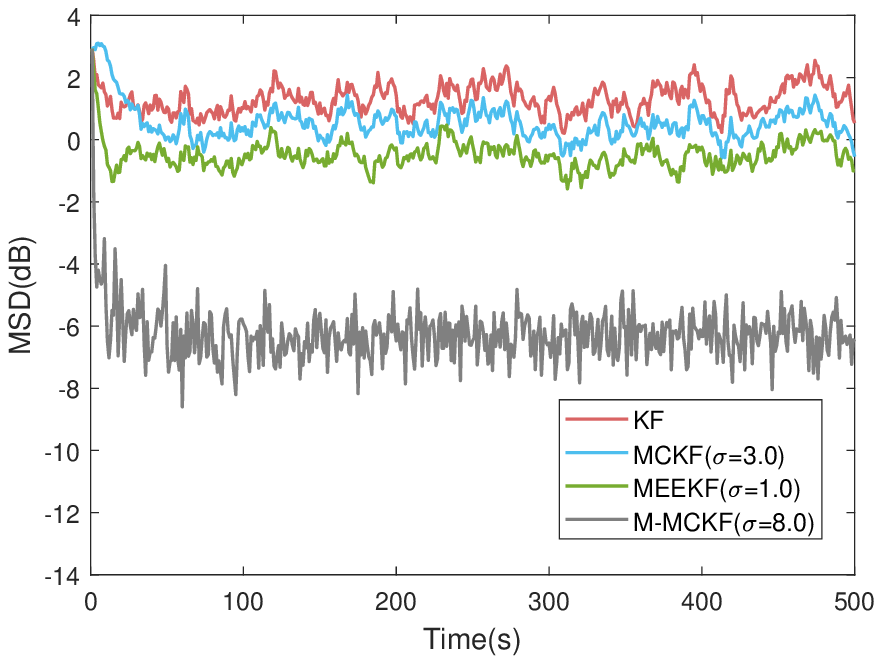}%
\label{mmckf}}
\hfil
\subfloat[]{\includegraphics[width=0.5\columnwidth]{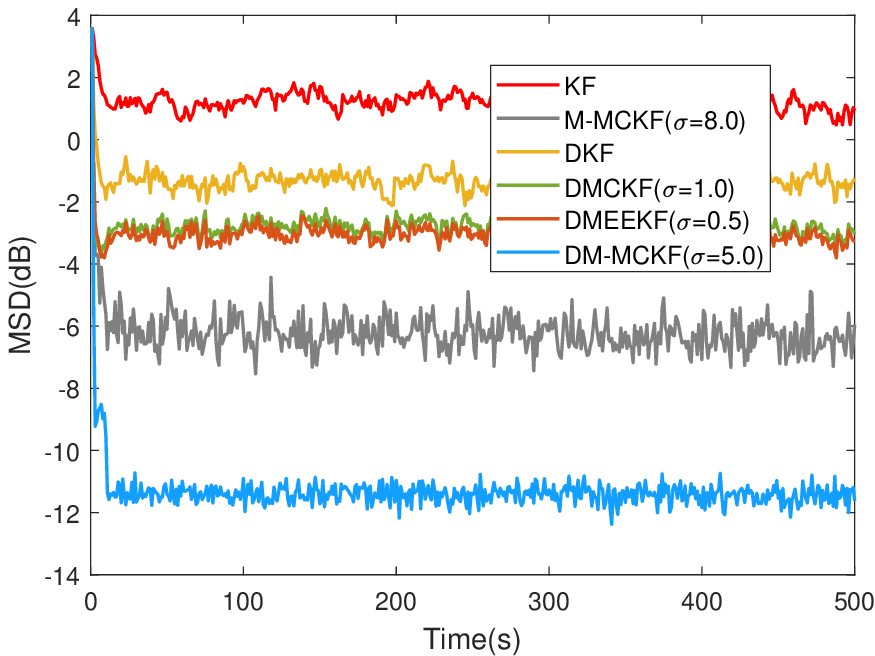}%
\label{dmmckf}}
\caption{The validation of the proposed M-MCKF and DM-MCKF algorithms. (a) Comparison of the performance of the M-MCKF algorithm. (b) Comparison of the performance of the DM-MCKF algorithm.}
\label{mcdmcxn}
\end{figure}

In this example, the MSD of the M-MCKF algorithm is compared with that of KF, MCKF \cite{chen2017maximum}, and R-MEEKF \cite{WANG2021107914}. The simulation findings are displayed in Fig.\ref{mmckf} and Table \ref{MSD_Different_noise}. Fig.\ref{mmckf} shows the MSD of the M-MCKF algorithm and competitor, as well as the parameters of these algorithms; the stable MSDs of different algorithms in different scenarios are presented in Table \ref{MSD_Different_noise}, where N/A represent a situation where it is not applicable. From the simulation results, one can obtain that the M-MCKF performs best in terms of stable MSD with non-Gaussian noise; and the performance of the M-MCKF is comparable to that of the conventional KF algorithm. These findings demonstrate the robustness of the M-MCKF in both Gaussian and non-Gaussian noise environments.

\begin{figure}
\centerline{\includegraphics[width=0.7\columnwidth]{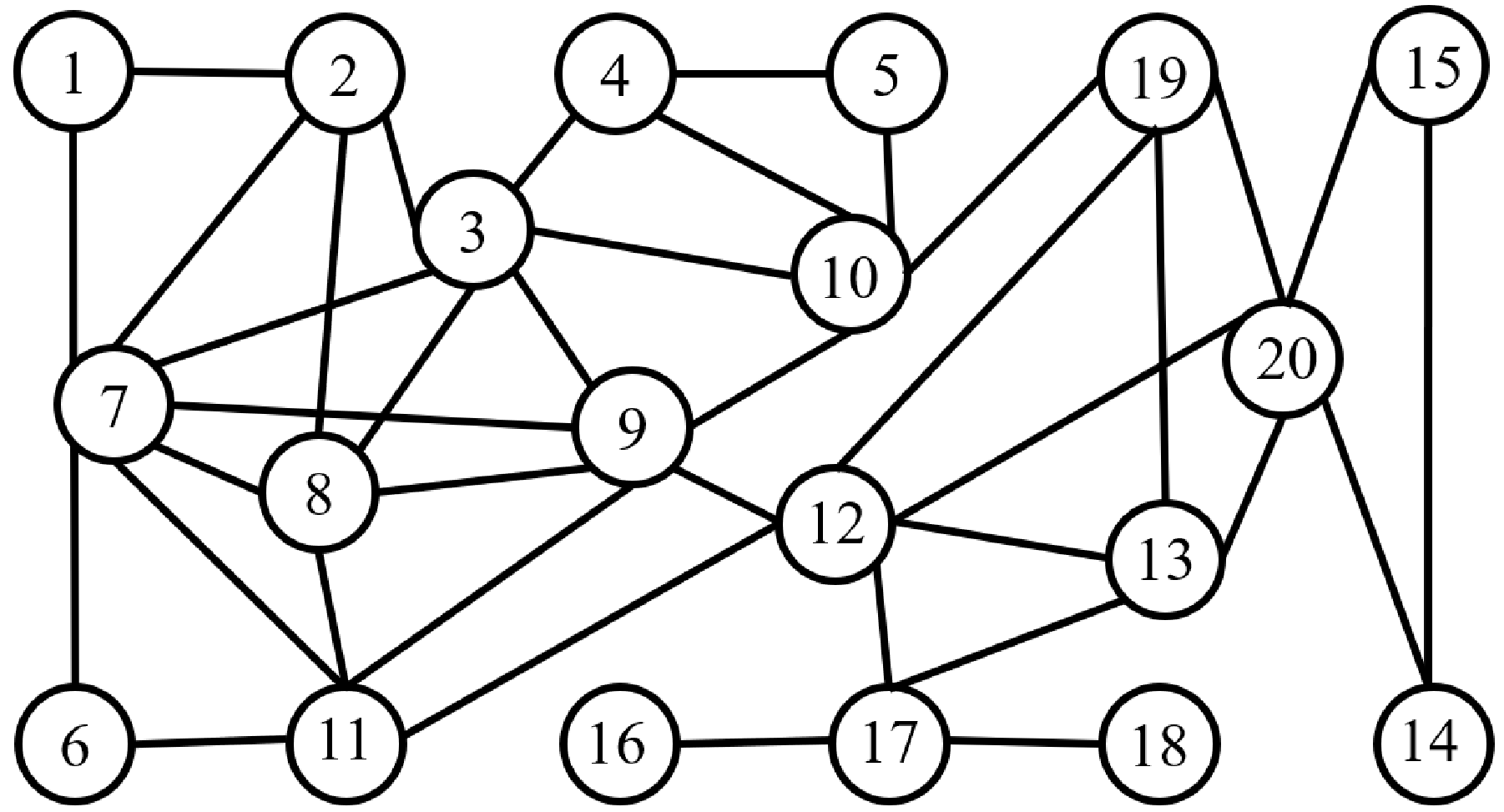}}
\caption{Network topology of the example WSNs.}\label{topuWSN}
\end{figure}

The performance of the distributed algorithm based on the proposed M-MCKF algorithm (DM-MCKF) is compared with that of DKF, D-MCKF \cite{wang2019distributed}, and DMEEKF \cite{feng2023distributed}. The state estimation performance is demonstrated by a classic WSNs example \cite{5411741} with 20 sensor nodes, and the nodes and connections are illustrated in Fig. \ref{topuWSN}. All simulation conditions are set in the same way as in the simulation above. The simulation results and parameter settings are shown in Fig. \ref{dmmckf} and Table \ref{MSD_Different_noisedmmckf}. Fig. \ref{dmmckf} illustrates the instantaneous MSD of the DM-MCKF algorithm and several competitors in the second scenario, and the node number is $i=7$; Table \ref{MSD_Different_noisedmmckf} presents the steady-state MSD of the DM-MCKF, DMCKF, and DKF at nodes 7, 13, and 16, respectively. From simulation results, one can obtain that 1) the DM-MCKF algorithm performs best with mixed-Gaussian, Laplace, and $\alpha$-stable noise, and its performance is comparable to that of the DKF with Gaussian noise in terms of stead-state MSD, which demonstrates the robustness of the DM-MCKF algorithm; 2) for the same algorithm, the method performs better as more nodes are added that are next to it.   

\begin{table*}[htbp]
\centering
\caption{The steady-state MSD (dB) comparison in different scenarios.}\label{MSD_Different_noise}
\begin{tabular}{lllllllll}
\hline
      & \multicolumn{2}{l}{First scenario} & \multicolumn{2}{l}{Second scenario} & \multicolumn{2}{l}{Third scenario} & \multicolumn{2}{l}{Fourth scenario} \\
            & MSD               & $\sigma$          & MSD               & $\sigma$           & MSD               & $\sigma$          & MSD               & $\sigma$           \\ \hline
KF          & \textbf{-7.66}   & N/A            & 1.24            & N/A             & 1.83           & N/A           & 5.91            & N/A             \\
MCKF        & -7.63           & 30.0           & 0.34            & 3.0             & 1.59           & 4.0           & 4.68            & 2.5             \\
R-MEEKF     & -7.64           & 15.0           & 0.48            & 1.0             & 1.55           & 2.0           & 4.15            & 1.2             \\
M-MCKF      & -7.65           & 8.0            & \textbf{-6.23}   & 8.0             & \textbf{1.36}   & 3.5           & \textbf{2.19}    & 3.0               \\ \hline
\end{tabular}
\end{table*}

\begin{table*}[htbp]
\centering
\caption{The steady-state MSD (dB) of the distributed algorithms in different scenarios.}\label{MSD_Different_noisedmmckf}
\begin{tabular}{lllllllll}
\hline
   & \multicolumn{2}{l}{First scenario} & \multicolumn{2}{l}{Second scenario} & \multicolumn{2}{l}{Third scenario} & \multicolumn{2}{l}{Fourth scenario} \\
              & MSD              & $\sigma$           & MSD               & $\sigma$           & MSD              & $\sigma$           & MSD              & $\sigma$            \\ \hline
KF            & -7.66    & N/A    & 1.24              & N/A             & 0.34             & N/A             & 5.90             & N/A              \\
M-MCKF        & -7.65    & 8.0             & -6.23             & 8.0             & 1.36             & 3.5             & 2.19             & 3.0              \\
DKF($i=16$)     & -9.08    & N/A             & 0.52           & N/A             & 0.91             & N/A             & N/A              & N/A              \\
DKF($i=13$)     & -10.92   & N/A             & -0.61             & N/A             & -0.29            & N/A             & N/A              & N/A              \\
DKF($i=7$)      & \textbf{-12.0} & N/A             & -1.38             & N/A             & -0.81            & N/A             & N/A              & N/A              \\
DMCKF($i=16$)   & -9.08          & 10.0            & -0.94           & 1.0             & 0.57             & 2.0             & 3.22             & 2.0              \\
DMCKF($i=13$)   & -10.92         & 10.0            & -2.18             & 1.0             & -0.49            & 2.0             & 2.08             & 2.0              \\
DMCKF($i=7$)    & -12.0          & 10.0            & -2.81             & 1.0             & -1.07            & 2.0             & 1.82             & 2.0              \\
DM-MCKF($i=16$) & -9.06          & 5.0             & -8.21             & 5.0             & 0.36             & 3.0             & 1.67             & 2.5              \\
DM-MCKF($i=13$) & -10.8          & 5.0             & -10.36            & 5.0             & -0.67            & 3.0             & 1.18             & 2.5              \\
DM-MCKF($i=7$)  & -11.89         & 5.0             & \textbf{-11.40}            & 5.0             & \textbf{-1.21}            & 3.0             & \textbf{0.91}             & 2.5              \\ \hline
\end{tabular}
\end{table*}

\subsection{Performance verification of the proposed algorithms considering packet drop}
\begin{figure}[!t]
\centering
\subfloat[]{\includegraphics[width=0.5\columnwidth]{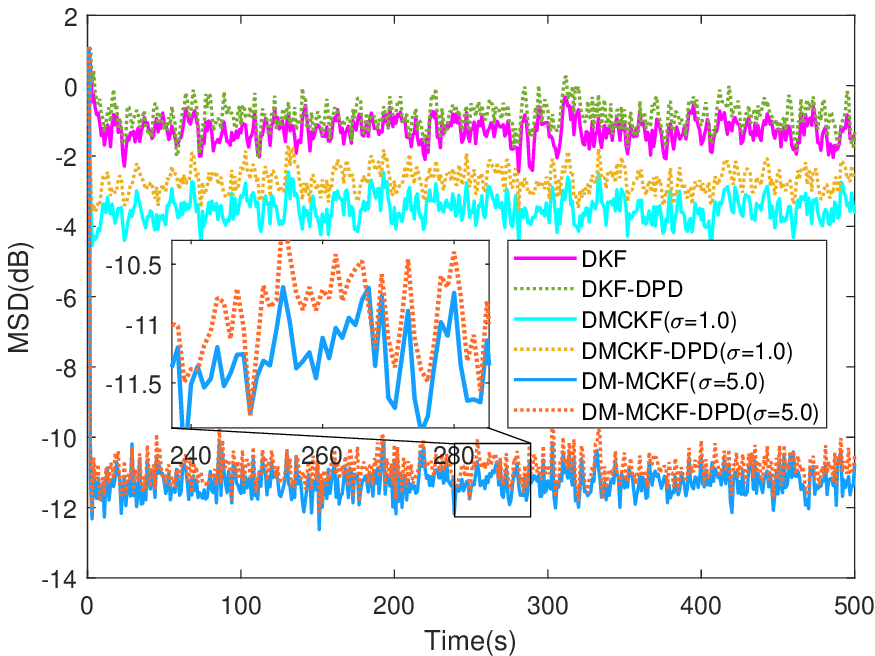}%
\label{mmckfdpd}}
\hfil
\subfloat[]{\includegraphics[width=0.5\columnwidth]{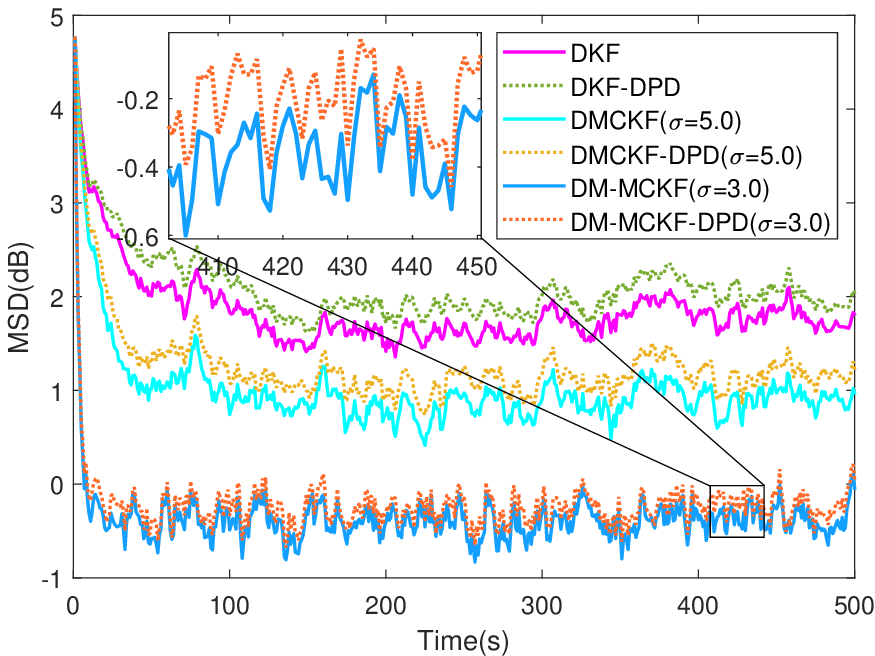}%
\label{dmmckfdpd}}
\caption{The performance of the DM-MCKF-DPD algorithm in different scenarios. (a) The convergence curves in the second scenario. (b) The convergence curves in the third scenario.}
\label{mcdmcxndpd}
\end{figure}

In addition, the state estimation performance is verified by the DM-MCKF-DPD, DMCKF-DPD, and stationary DKF algorithm \cite{8409298} when applying them to a classic WSNs example \cite{5411741} under data packet drop and impulsive noise conditions. It is assumed that the probability that all nodes can receive information from neighboring nodes is $p_k^{i,j} = 0.8$, and the kernelwidths are shown in Fig. \ref{mcdmcxndpd}. The performance of the proposed algorithms is demonstrated with different nodes and different scenarios, and the corresponding results are shown in Fig. \ref{mcdmcxndpd} and Table. Fig. \ref{mmckfdpd} shows the convergence curves of MSD in the second scenario and the node number is $i=7$; Fig. \ref{dmmckfdpd} shows the convergence curves of MSD in the third scenario and the node number $i=13$. From Fig. \ref{mcdmcxndpd}, one can obtain that 1) the proposed DM-MCKF-DPD and DMCKF-DPD algorithms perform better than the DKF-DPD algorithm with non-Gaussian noise; 2) DM-MCKF-DPD algorithm performs better than DMCKF-DPD algorithm.

In addition, the C-DM-MCKF-DPD algorithm is verified in this part. The kernelwidth is set to 5.0 and the consensus coefficient is set as $\eta=0.05$. The convergence curve of MSD of the $17$th node in the second scenario is shown in Fig. \ref{cdmcdmcxn}. From the simulation result, one can obtain that the consensus scheme ameliorates the impact of communication data loss to some extent.

\subsection{Parameters discussion}
The influence of $\sigma$ on the performance of the M-MCKF and DM-MCKF algorithms is studied in this part. The influence of the consensus coefficient $\eta$ on the performance of the consensus DM-MCKF-DPD algorithm is investigated. The conclusions obtained can be used to guide the choice of the proposed algorithms. 

\begin{figure}[!t]
\centering
\subfloat[]{\includegraphics[width=0.5\columnwidth]{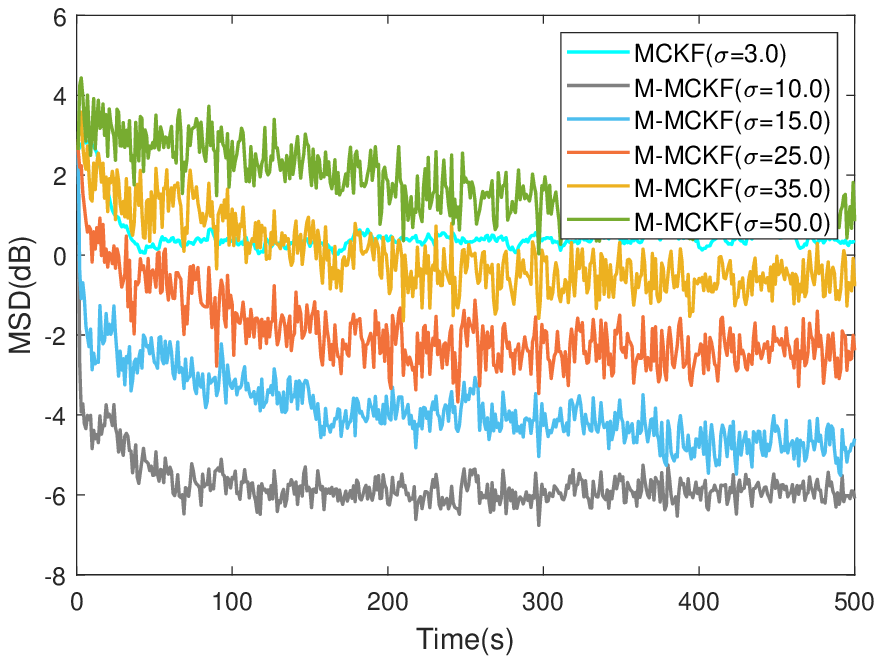}%
\label{mmckf_sigma}}
\hfil
\subfloat[]{\includegraphics[width=0.5\columnwidth]{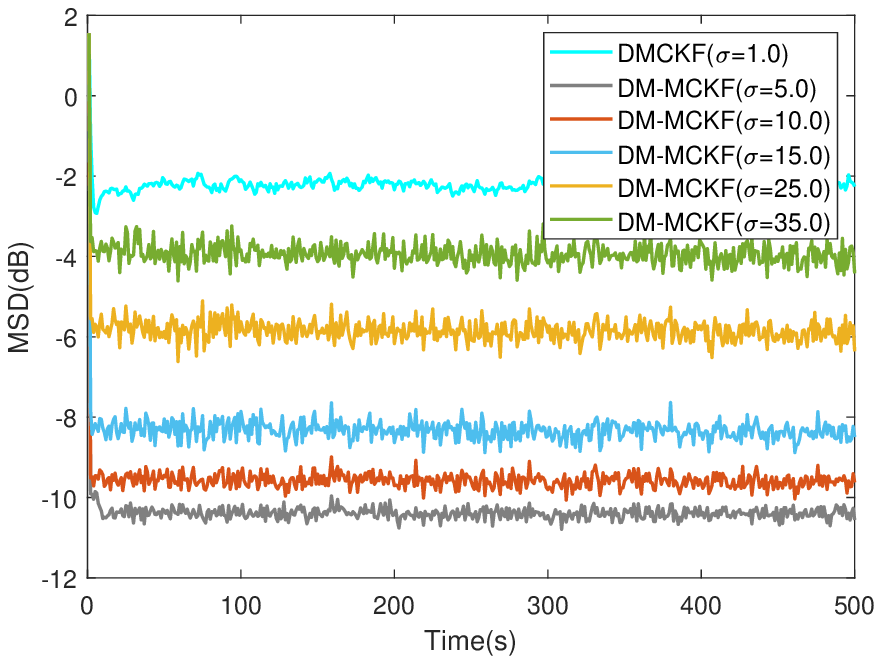}%
\label{dmmckf_sigma}}
\caption{The performance analysis of the proposed algorithms with different kernelwidths. (a) The M-MCKF algorithm with different $\sigma$ in the second scenario. (b) The DM-MCKF algorithm with different $\sigma$.}
\label{dmcdmcxn_sigma}
\end{figure}

The parameter $\sigma$ is set as $\sigma {\text{ = 3}}{\text{.0, 5}}{\text{.0, 8}}{\text{.0, 10}}{\text{.0, 15}}{\text{.0, 20}}{\text{.0, 35}}{\text{.0, 45}}{\text{.0}}$, and the initial states of the system are the same as in \eqref{inistate}. The simulation results are presented in Fig. \ref{dmcdmcxn_sigma}, Table \ref{tabmmckfsigma}, and Table \ref{tabdmmckfsigma}. Fig. \ref{dmcdmcxn_sigma} displays the convergence curves of the M-MCKF and DM-MCKF algorithms with different $\sigma$, respectively. The steady-state MSD, in the four scenarios, of the M-MCKF algorithm with different $\sigma$ is presented in Table \ref{tabmmckfsigma}. The steady-state MSD, in the four scenarios, of the DM-MCKF algorithm with different $\sigma$ and nodes is shown in Table \ref{tabdmmckfsigma}. From Fig. \ref{mmckf_sigma} and Table \ref{tabmmckfsigma}, one can obtain that the performance of the M-MCKF algorithm improves with the increasing $\sigma$ with Gaussian measurement noise, the it is optimal for mixed-Gaussian, Laplace, and $\alpha$-stable noise when $\sigma$ are around 8.0, 3.0, 3.0. From Fig. \ref{dmmckf_sigma} and Table \ref{tabdmmckfsigma}, we can infer that the proposed DM-MCKF algorithm performs best with above the non-Gaussian noise when $\sigma$ is around 3.0. In addition, it is clear that the higher the number of neighboring nodes the better the performance of the DM-MCKF algorithm.

\begin{figure}[!t]
\centering
\subfloat[]{\includegraphics[width=0.5\columnwidth]{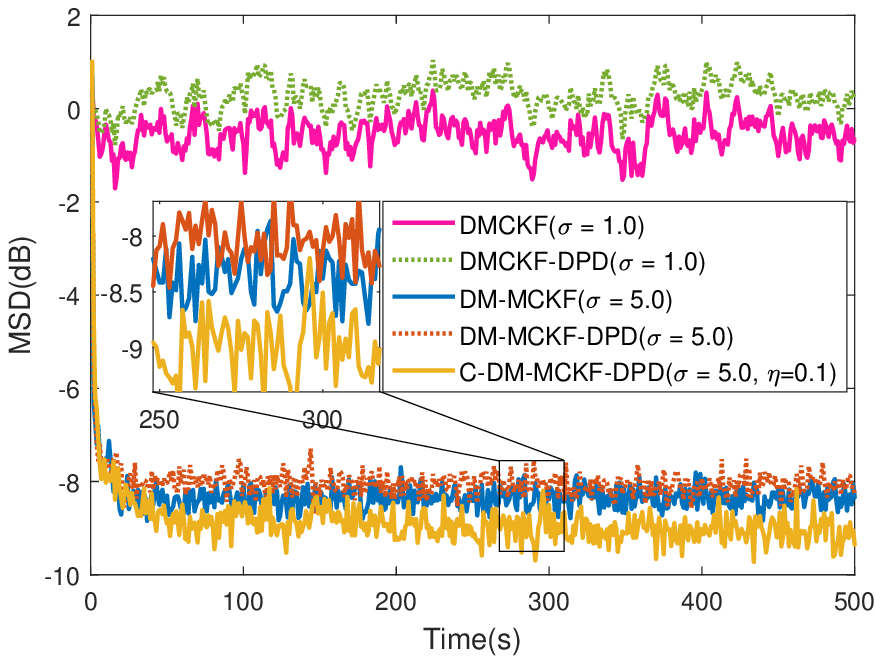}%
\label{cdmcdmcxn}}
\hfil
\subfloat[]{\includegraphics[width=0.5\columnwidth]{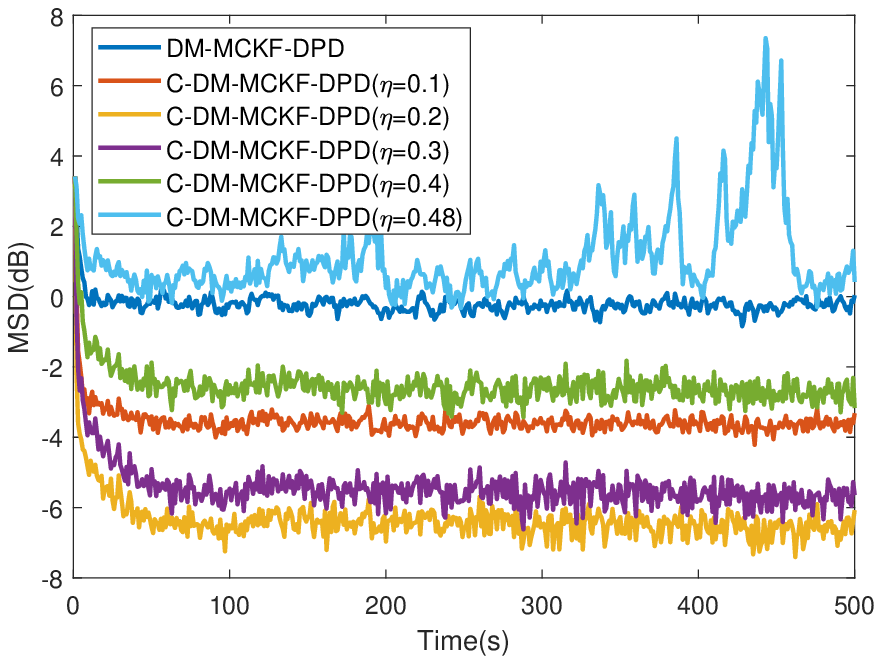}%
\label{cdmcdmcxnparaeta}}
\caption{The performance comparison and analysis of consensus strategy. (a) The performance comparison of consensus strategy. (b) The consensus strategy with different $\eta$.}
\label{cdmcdmcxn_eta}
\end{figure}

\begin{figure}[!t]
\centering
\subfloat[]{\includegraphics[width=0.5\columnwidth]{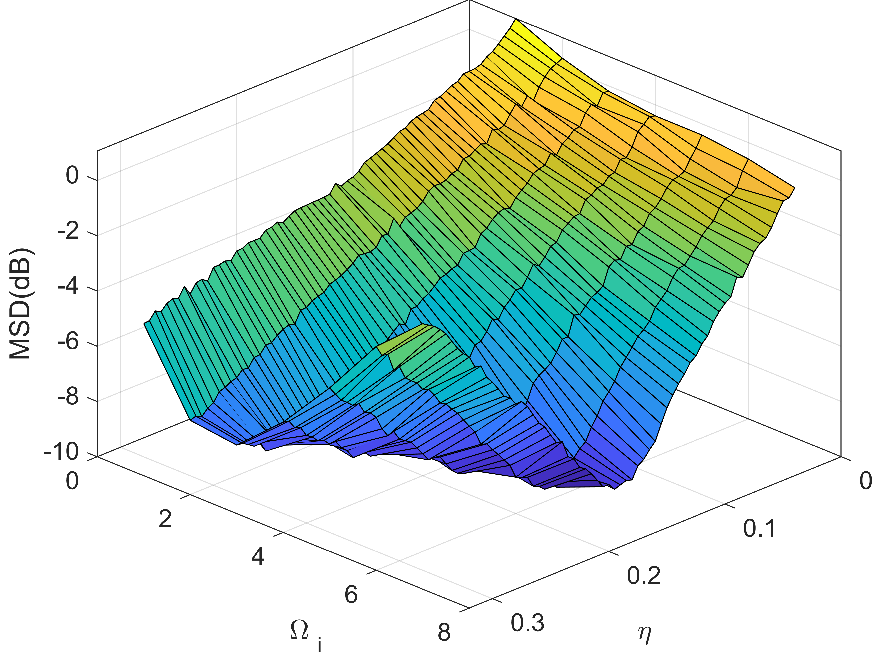}%
\label{MSD_node_eta_laplace}}
\hfil
\subfloat[]{\includegraphics[width=0.5\columnwidth]{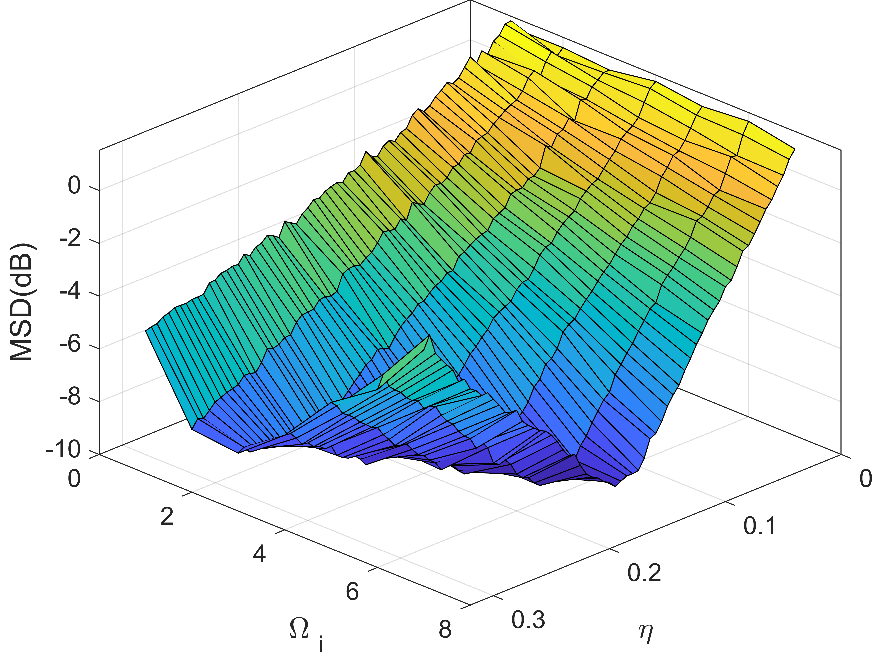}%
\label{MSD_node_eta_Alpha}}
\caption{The performance surfaces with respect to parameter $\eta$ and ${\Omega _i}$. (a) The performance surface in the third scenario. (b) The performance surface in the fourth scenario.}
\label{msd_eat_node}
\end{figure}

The parameter $\eta$ is set as $\eta {\text{ = }}0.1,{\text{ 0}}{\text{.2, 0}}{\text{.4, 0}}{\text{.48}}$, the kernelwidth of DM-MCKF-DPD and C-DM-MCKF-DPD are 3, and $p_k^{i,j} = 0.9$. In the third scenario, the convergence curves of MSD of the $17$th node are displayed in Fig. \ref{cdmcdmcxnparaeta}. In addition, the performance surfaces for the influence of the number of neighboring nodes ${\Omega _i}$ and the consensus coefficient $\eta$ on the performance of the C-DM-MCKF-DPD algorithm are shown in Fig. \ref{msd_eat_node}. From the simulation presented in Fig. \ref{cdmcdmcxnparaeta}, one can obtain that the optimal consensus coefficient is around 0.2 for the $17$th node in the third scenario. From Fig. \ref{msd_eat_node}, one can obtain that 1) when the number of neighbors of a node ${\Omega _i}$ is small ($0 < {\Omega _i} \leqslant 4$), choosing a larger consensus coefficient ($0.2 \leqslant \eta  \leqslant 0.35$) can improve the performance of the consensus algorithm; 2) when the number of neighbors of a node ${\Omega _i}$ is large ($4 < {\Omega _i} \leqslant 7$), choosing a smaller consensus coefficient ($0.1 \leqslant \eta  \leqslant 0.2$) can improve the performance of the consensus algorithm.

\begin{table*}
\centering
\caption{The steady-state MSD (dB) of M-MCKF algorithm with different $\sigma$.}\label{tabmmckfsigma}
\begin{tabular}{llllllllll}
\hline
                & $\sigma=2$ & $\sigma=3$     & $\sigma=5$     & $\sigma=8$     & $\sigma=10$    & $\sigma=15$    & $\sigma=20$    & $\sigma=35$    & $\sigma=45$    \\ \hline
First scenario  & 8.14  & 5.99  & -7.52  & -7.62 & -7.63 & -7.64 & -7.64 & -7.64 & \textbf{-7.64} \\
Second scenario & 20.98 & 14.56 & 10.32  & \textbf{-6.35} & -5.90 & -4.71 & -3.47 & -0.49 & 0.70  \\
Third scenario  & 1.81  & \textbf{1.35}  & 1.48  & 1.77  & 1.86  & 1.97  & 2.18  & 2.24  & 2.26  \\
Fourth scenario & 2.68  & \textbf{2.12}  & 2.23  & 2.42  & 2.59  & 2.92  & 3.12  & 3.74  & 3.96  \\ \hline
\end{tabular}
\end{table*}

\begin{table*}
\centering
\caption{The steady-state MSD (dB) of the DM-MCKF algorithm with different $\sigma$.}\label{tabdmmckfsigma}
\begin{tabular}{llllllllll}
\hline
                      & $\sigma=2$ & $\sigma=3$      & $\sigma=5$      & $\sigma=8$      & $\sigma=10$     & $\sigma=15$     & $\sigma=20$     & $\sigma=35$     & $\sigma=45$     \\ \hline
First scenario(i=16)  &  4.82 & -8.94  & -8.99  & -9.00  & -9.00  & -9.00  & -9.00  & -9.00  & \textbf{-9.00}  \\
First scenario(i=13)  &  2.63 & -10.80 & -10.84 & -10.85 & -10.85 & -10.85 & -10.85 & -10.85 & \textbf{-10.85} \\
First scenario(i=7)   & -1.04 & -11.78 & -11.89 & -11.99 & -11.99 & -11.99 & -11.99 & -11.99 & \textbf{-11.99} \\
Second scenario(i=16) & 14.65 & \textbf{-8.40} & -8.23  & -7.49  & -6.98  & -5.51  & -4.33  & -1.25  & -0.10  \\
Second scenario(i=13) & 10.37 & \textbf{-10.45} & -10.37 & -9.98  & -9.56  & -8.30  & -6.94  & -3.89  & -2.54  \\
Second scenario(i=7)  &  9.92 & \textbf{-11.43} & -11.42 & -11.02 & -10.62 & -9.38  & -8.14  & -5.10  & -3.70  \\
Third scenario (i=16) &  1.34 & \textbf{0.53}   & 0.66   & 0.88   & 1.06   & 1.21   & 1.25   & 1.33   & 1.36   \\
Third scenario(i=13)  &  0.40 & \textbf{-0.67}  & -0.43  & -0.07  & 0.01   & 0.21   & 0.26   & 0.31   & 0.34   \\
Third scenario(i=7)   & -0.16 & \textbf{-1.21}  & -1.01  & -0.78  & -0.55  & -0.46  & -0.37  & -0.32  & 0.22   \\
Fourth scenario(i=16) & 2.07 & \textbf{1.72}   & 1.89   & 2.12   & 2.13   & 2.64   & 2.68   & 3.03   & 3.48   \\
Fourth scenario(i=13) & 1.62 & \textbf{1.11}   & 1.23   & 1.41   & 1.42   & 1.74   & 1.88   & 2.36   & 2.62   \\
Fourth scenario(i=7)  & 1.30 & \textbf{0.91}   & 1.01   & 1.08   & 1.22   & 1.43   & 1.62   & 2.04   & 2.28   \\ \hline
\end{tabular}
\end{table*}

\section{Conclusion}\label{section:Conclusion}
The distributed state estimation over wireless sensor network with presence of non-Gaussian noise was concerned in this paper. By proposed a generalized packet drop model, the process of packet loss due to DoS attacks and energy limitations is described. A modified maximum correntropy KF algorithm was developed by analysing the advantages and disadvantages of existing algorithms, and it was extended to distributed M-MCKF algorithm. In addition, the distributed modified maximum correntropy Kalman filter incorporating the proposed generalized packet drop model was developed. The computational complexity of the DM-MCKF-DPD algorithm was demonstrated to be moderate compared to that of the conventional stationary DKF, and a sufficient condition to ensure the convergence of the fixed-point iterative algorithm was presented. Finally, simulations conducted with a 20-node WSNs demonstrated that the proposed DM-MCKF and DM-MCKF-DPD algorithms perform better than some existing algorithms. As future work, we intend to investigate the state estimation performance of a distributed Kalman filter based on the MEE criterion for WSNs under data packet drop conditions to further improve estimated performance with data packet drop and non-Gaussian noise.

\bibliographystyle{unsrt}
\bibliography{cas-refs}

\begin{IEEEbiography}
[{\includegraphics[width=1in,height=1.25in,clip,keepaspectratio]{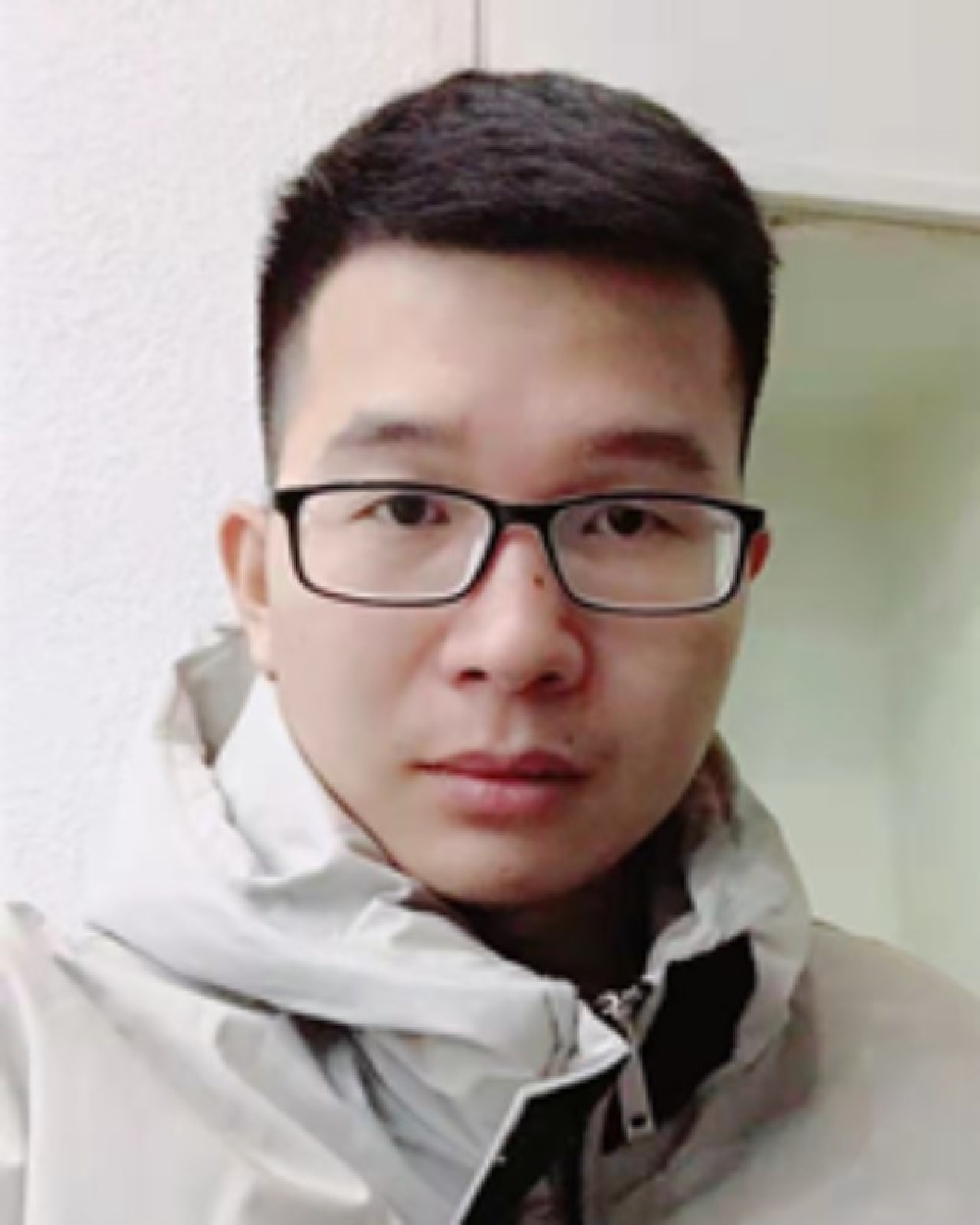}}] 
{Jiacheng He}{\space} received the B.S. degree in mechanical engineering from University of Electronic Science and Technology of China, Chengdu, China, in 2020. He is currently pursuing a Ph.D. degree in the School of Mechanical and Electrical Engineering, University of Electronic Science and Technology of China, Chengdu, China. His current research interests include information-theoretic learning, signal processing, and target tracking.
\end{IEEEbiography}

\begin{IEEEbiography}
[{\includegraphics[width=1in,height=1.25in,clip,keepaspectratio]{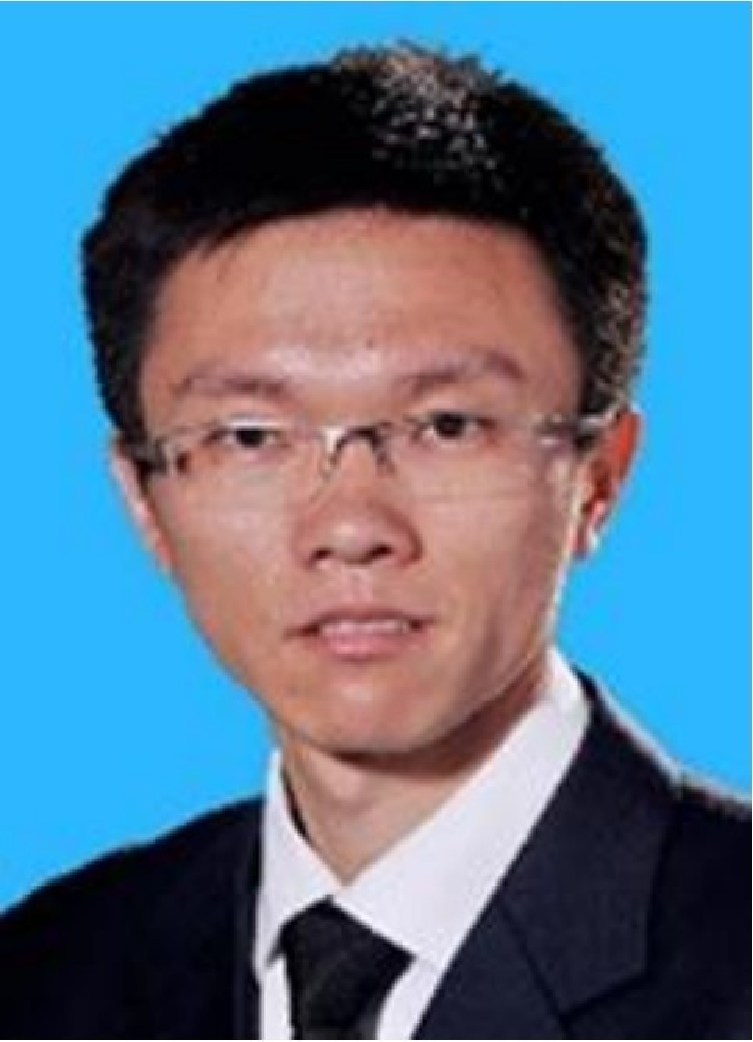}}] 
{Bei Peng}{\space} received the B.S. degree in mechanical engineering from Beihang University, Beijing, China, in 1999, and the M.S. and Ph.D. degrees in mechanical engineering from Northwestern University, Evanston, IL, USA, in 2003 and 2008, respectively. He is currently a Full Professor of Mechanical Engineering with the University of Electronic Science and Technology of China, Chengdu, China. He holds 30 authorized patents. He has served as a PI or a CoPI for more than ten research projects, including the National Science Foundation of China. His research interests mainly include intelligent manufacturing systems, robotics, and its applications.
\end{IEEEbiography}

\begin{IEEEbiography}
[{\includegraphics[width=1in,height=1.25in,clip,keepaspectratio]{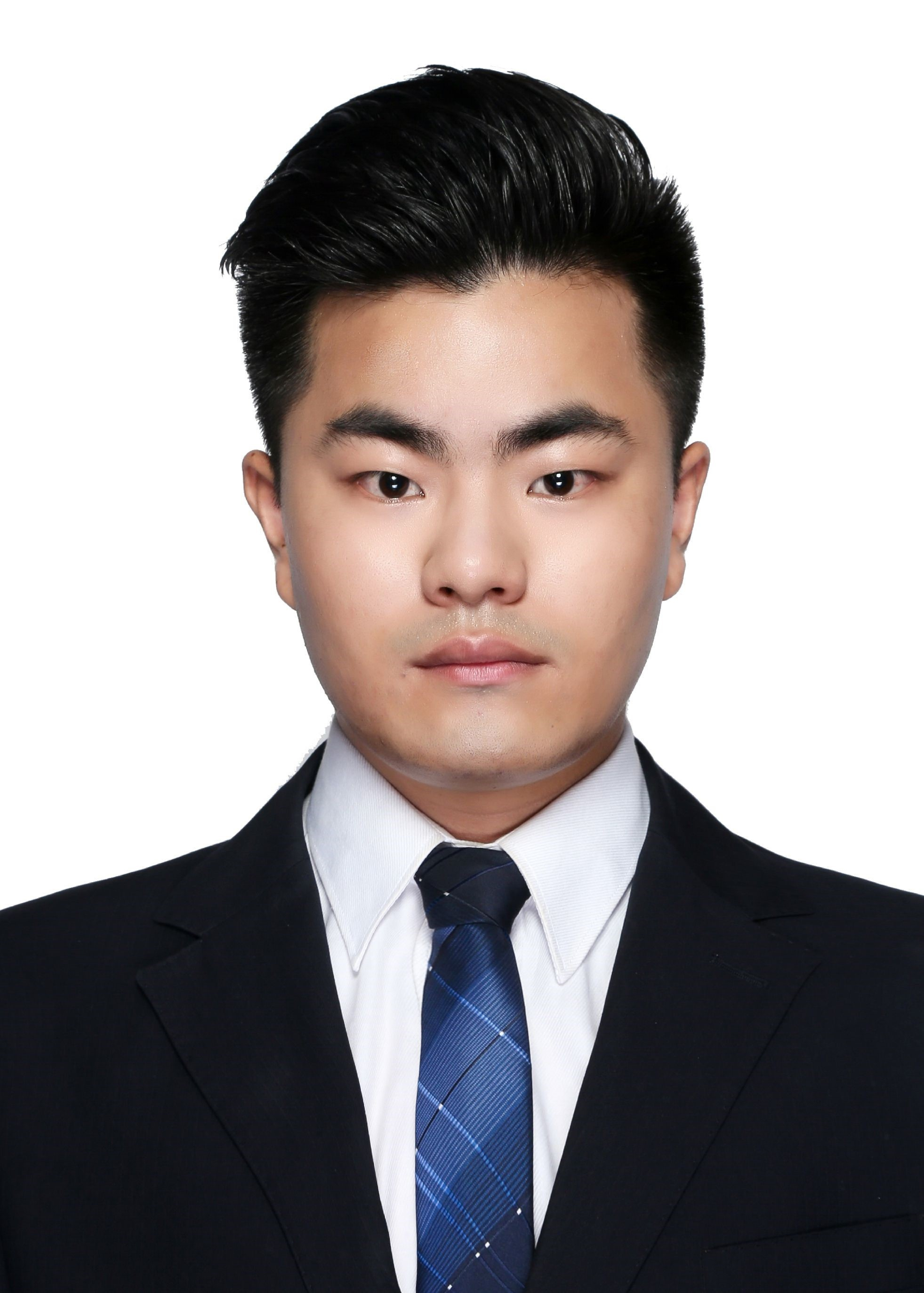}}] 
{Zhenyu Feng}{\space} was born in Anhui, China. He received the B.E. and MA.Sc degree from the University of Electronic Science and Technology of China, Chengdu, China, in 2014 and 2018, respectively. He is currently working toward the Ph.D. degree in the school of Mechanical and Electrical Engineering at University of Electronic Science and Technology of China. He has over five years of industrial experience in designing and implementing for applications in multi-agent sensor network systems. His current research interests lie in the field of multi-agent systems, underwater unmanned vehicle swarm intelligence, as well as distributed communication technology. 
\end{IEEEbiography}

\begin{IEEEbiography}
[{\includegraphics[width=1in,height=1.25in,clip,keepaspectratio]{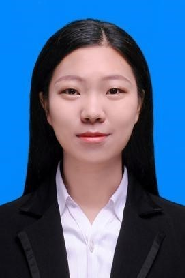}}] 
{Xuemei Mao}{\space} received the B.S. degree in mechanical design, manufacturing, and automation from Xidian University, Xian, China, in 2020. She is currently pursuing the Ph.D. degree in mechanical engineering with School of Mechanical and Electrical Engineering, University of Electronic Science and Technology of China, Chengdu, China. Her current research interests include signal processing, adaptive filtering, and target tracking.
\end{IEEEbiography}

\begin{IEEEbiography}
[{\includegraphics[width=1in,height=1.25in,clip,keepaspectratio]{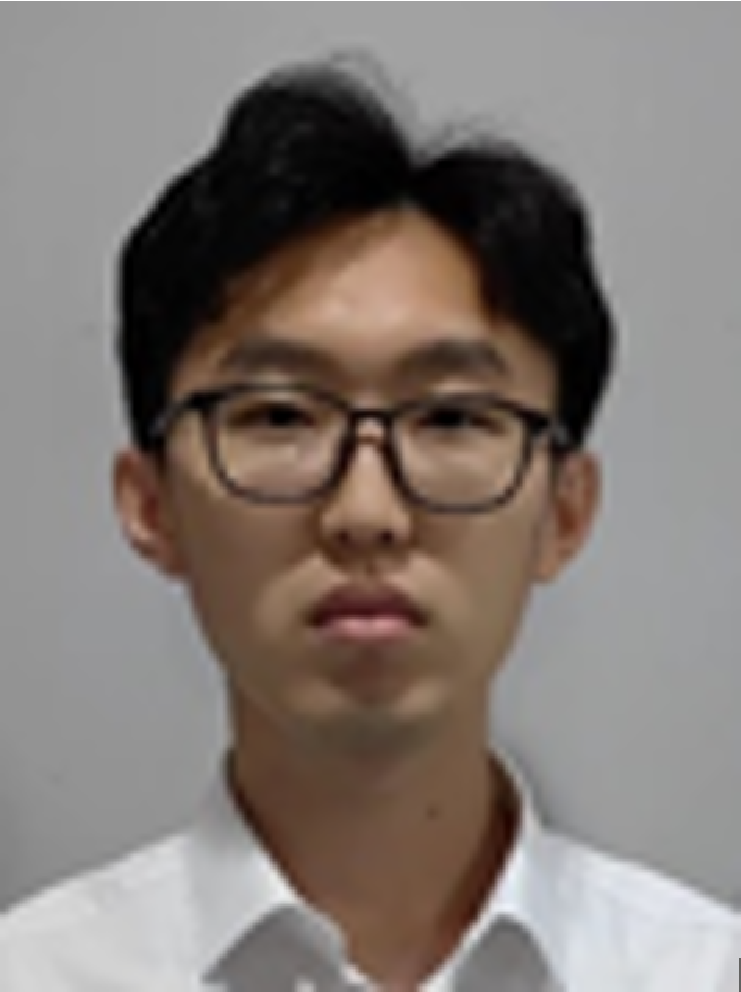}}] 
{Song Gao}{\space} received the B.S. degree in mechanical design, manufacturing and automation from University of Electronic Science and Technology of China (UESTC), in 2020. He is currently pursuing the Ph.D. degree in mechanical engineering with School of Mechanical and Electrical Engineering of UESTC, Chengdu, China. His current research interests include multi-agent systems, consensus control, signal processing.
\end{IEEEbiography}

\begin{IEEEbiography}
[{\includegraphics[width=1in,height=1.25in,clip,keepaspectratio]{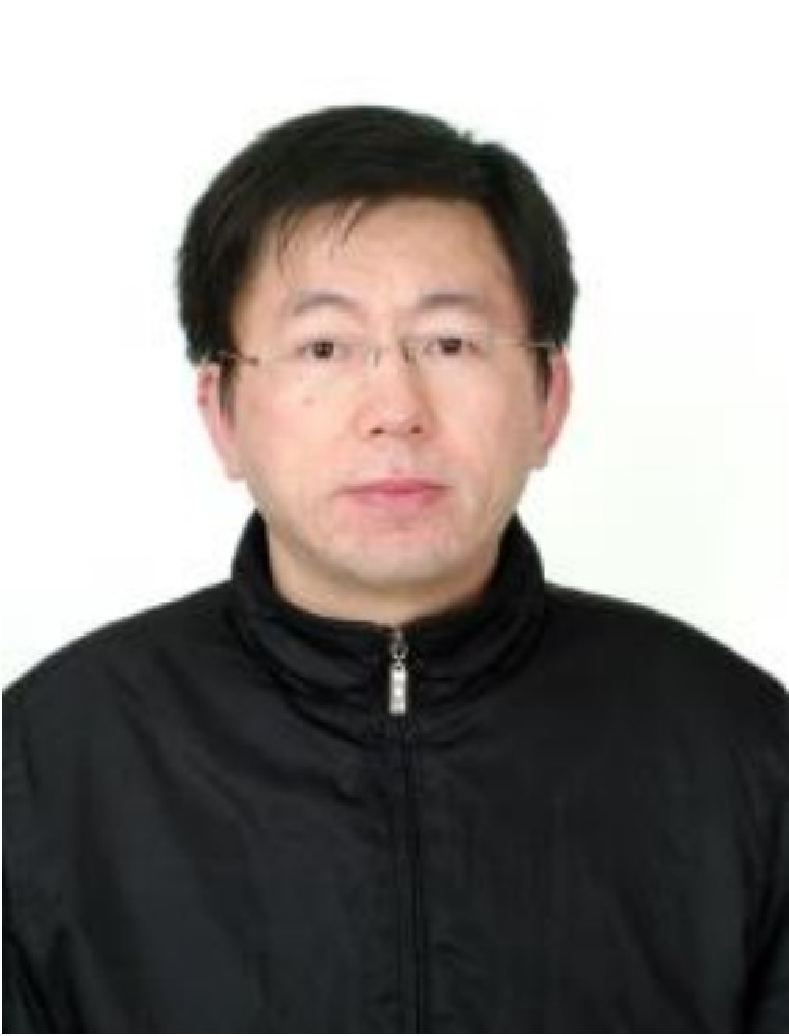}}] 
{Gang Wang}{\space} received the B.E. degree in Communication Engineering and the Ph.D. degree in Biomedical Engineering from University of Electronic Science and Technology of China, Chengdu, China, in 1999 and 2008, respectively. In 2009, he joined the School of Information and Communication Engineering, University of Electronic Science and Technology of China, China, where he is currently an Associate Professor. His current research interests include signal processing and intelligent systems.
\end{IEEEbiography}
\end{CJK}
\end{document}